\documentclass[onecolumn]{elsart3p}

\usepackage{amssymb}
\usepackage{graphicx,psfrag}
\usepackage{amsfonts}
\usepackage{amssymb}
\usepackage{amsmath}
\usepackage{epsfig, color}
\usepackage{setspace}
\usepackage{epstopdf}
\usepackage{bm}

\usepackage[english,francais]{babel}

\newtheorem{e-proposition}[theorem]{Proposition}

\newtheorem{e-definition}[theorem]{Definition\rm}

\renewcommand{\theequation}{\arabic{equation}}
\def\va{{\vec a}}
\def\vk{{\vec k}}
\def\vr{{\vec r}}
\def\vq{{\vec q}}
\def\vp{{\vec p}}
\def\vb{{\vec b}}
\def\vd{{\vec d}}
\def\vl{{\vec l}}

\def\vK{{\vec K}}

\def\vB{{\vec B}}

\def\vp{{\vec p}}

\def\ve{{\vec e}}

\def\ve{{\vec e}}

\def\dhat{{\hat {\vec d}}}

\def\og{\leavevmode\raise.3ex\hbox{$\scriptscriptstyle\langle\!\langle$~}}
\def\fg{\leavevmode\raise.3ex\hbox{~$\!\scriptscriptstyle\,\rangle\!\rangle$}}

\journal{CR Physique}

\begin{document}
\centerline{Title of the dossier/Titre du dossier}
\begin{frontmatter}


\selectlanguage{english}
\title{Introduction to Dirac materials and topological insulators}


\selectlanguage{english}
\author[authorlabel1,authorlabel2]{J\'er\^ome Cayssol},
\ead{author.name1@email.address1}

\address[authorlabel1]{Max-Planck-Institut f\"ur Physik komplexer Systeme, N\"othnitzer Str. 38, 01187 Dresden, Germany}
\address[authorlabel2]{LOMA (UMR-5798), CNRS and University Bordeaux 1, F-33045 Talence, France}

\begin{abstract}
We present a short pedagogical introduction to the physics of Dirac materials, restricted to graphene and two-dimensional topological insulators. We start with a brief reminder of the Dirac and Weyl equations in the particle physics context. Turning to condensed matter systems, semimetallic graphene and various Dirac insulators are introduced, including the Haldane and the Kane-Mele topological insulators. We also discuss briefly experimental realizations in materials with strong spin-orbit coupling.

\end{abstract}
\end{frontmatter}

\selectlanguage{english}
\section{Introduction}

The concepts of Dirac, Majorana and Weyl fields, commonly used to describe elementary particles (electrons, quarks, neutrinos,...), have recently entered the realm of condensed matter physics. In high-energy physics, those fields are the simplest building blocks to construct Lorentz invariant Lagrangians describing interacting particles within the standard model \cite{Zee:2010,Weinberg}. Historically, Dirac first introduced his famous wave equation in order to describe a free electron, satisfying the relativistic dispersion relation, $E^2=p^2 c^2 + m^2 c^4$, between its energy $E$, momentum $\vp$ and mass $m$ ($c$ being the velocity of light) \cite{Dirac:1928}. This equation has been built to be compatible with both single particle quantum mechanics and special relativity. Dirac further pointed out the difficulties raised by the existence of negative energy solutions of relativistic wave equations (including his own equation \cite{Dirac:1928,Dirac:1930} and the Klein-Gordon equation \cite{Klein:1926}). Those negative energy solutions are related to a new type of particles, the antiparticles, which have the same mass as the electron but couple to the electromagnetic field with the opposite charge \cite{Dirac:1930}. The positron, the antiparticle of the electron, was discovered soon after this theoretical prediction. For a massive particle in our familiar 3+1 space-time continuum, the Dirac wave function is a 4-component spinor which describes two spin one-half particles (the particle and its particle). A massless spin-one half particle is described by a two-component Weyl spinor \cite{Weyl:1929}. In contrast to the Dirac equation, the Weyl equation breaks space inversion because it can be written separately for left-handed (and right-handed) Weyl fermions. Finally let us mention that the Majorana field is a massive field, like the Dirac one, but describing a particle identical to its antiparticle (unlike the Dirac case) \cite{Majorana:1937,Pal:2010}. Hence the Majorana field has only two independent complex components \cite{Zee:2010,Weinberg}. 

\medskip
This paper is devoted to "Dirac materials", namely to lattice systems where the excitations are described by relativistic Dirac or Weyl equations. These materials are usually narrow (or zero) gap semiconductors where two (or more) bands get strongly coupled near a level-crossing. Due to the presence of the lattice, electrons are described by Bloch states indexed by a quasi-momentum $\vp$, and their energies $E(\vp)$ are periodic over the BZ. This implies that the Weyl or Dirac equations (and the corresponding dispersion relations) cannot be satisfied globally over the whole BZ, but only locally. The most celebrated "Dirac material" is graphene, the two-dimensional monolayer of carbon atoms, where massless Weyl excitations emerge near two isolated points of the reciprocal space. Graphene has two spin-degenerated Dirac cones (or equivalently 4 non degenerated Dirac cones).

\medskip

Meanwhile a new class of band insulators, the so-called topological insulators (TIs), has been discovered  \cite{KaneRMP:2011,QiRMP:2011,KonigJPSJ:2008,QiPhysToday:2010,Bernevig:2013}. The 2D topological insulator, also called Quantum Spin Hall (QSH) state, is distinguished from ordinary band insulators by the presence of a one-dimensional metal along its edge \cite{Kane:2005a,Kane:2005b} (Fig. 4 below). The nonchiral QSH edge states are also different from the chiral edge states of the Quantum Hall insulators or Chern insulators (Fig. 3), thereby providing a new class of one-dimensional (1D) conductors. Interestingly, the direction of the spin of the 1D charge carriers is tied to their direction of motion. Such conductors are protected from single-particle backscattering (and Anderson localization) by time-reversal symmetry $\mathcal T$. Interestingly, the QSH state has a three-dimensional (3D) generalization: the 3D TIs are (at least theoretically) insulating in the bulk, and exhibit topologically protected metallic states at their surfaces. Those two-dimensional (2D) surface states are characterized by a single (or an odd number of) non-degenerated Dirac cone(s). In those 2D Dirac surface states, the electron momentum is locked to the real spin in contrast to graphene where it is tied to the sublattice isospin.

\medskip

This short review is a pedagogical and (highly) non exhaustive introduction to Dirac materials, using graphene as a guideline. It is restricted to 2D Dirac materials and topological insulators in the absence of electron-electron interaction. We start with a short reminder of the Dirac \cite{Dirac:1928,Dirac:1930} and Weyl \cite{Weyl:1929} equations from the particle physics point of view. In Sec. \ref{SectionGraphene}, we turn to graphene as a lattice system whose band structure is described by Weyl-Dirac-like equations near some isolated points of the BZ. Sec. \ref{insulators} is devoted to the descriptions of various insulators obtained by gapping out the Dirac points of graphene with different mass terms. We also discuss recent experimental realizations in materials with strong spin-orbit coupling. Among these insulators, the Haldane state exhibits the quantized Hall effect and has topological features, like the existence of a topological invariant (Sec. \ref{chern}) and chiral edge states (Sec. \ref{SectionEdge}).

\section{Relativistic wave equations}
\label{SectionWave}
We briefly recall the Dirac and the Weyl equations for spin one-half fermions in the context of particle physics. The Dirac equation describes massive fermions \cite{Dirac:1928,Dirac:1930}, while massless particles obey the Weyl equation \cite{Weyl:1929}. Historically, these equations were introduced as relativistic wave equations for a single free particle. Nevertheless the existence of negative energy solutions requires to interpret these equations in the framework of (many-particle) quantum field theory \cite{Zee:2010,Weinberg}. The negative energy solutions correspond to a new type of particles, the antiparticles, which have the same mass as the particle but couple to the electromagnetic field with the opposite charge. 

\subsection{Dirac equation for massive fermions} 
Dirac introduced his famous equation as the simplest relativistic wave-equation describing a free electron in a 3+1 space-time continuum. He realized that the equation has to be first-order in the time-derivative:
\begin{equation}
i\hbar \frac{\partial \Psi}{\partial t} = H_D \Psi ,
\end{equation}
in order to ensure the interpretation of the wave function $\Psi(\vr,t)$ as a probability amplitude \cite{Dirac:1928}. Then Lorentz invariance implies that space-derivatives $\partial_\vr =(\partial_1,\partial_2,\partial_3)$ should appear at the same order (as the time derivative) suggesting the combination:
\begin{equation}
H_D = - i \hbar c \alpha_i \partial_i + \beta mc^2 ,
\end{equation}       
where $\alpha_i$ and $\beta$ are "some coefficients", and the summation over three space directions ($i=1,2,3$) is implied. The mass of the particle $m$, the speed of light $c$, and the reduced Planck constant $\hbar=h/2\pi$ are introduced in such a way that $\alpha_i$ and $\beta$ are dimensionless objects.

Applying twice the Hamiltonian $H_D$ on a plane wave $\Psi \simeq e^{i(\vp.\vr-Et)/\hbar}$ leads to the correct relativistic wave equation:
\begin{equation}
E^2 = p^2 c^2 + m^2 c^4,
\end{equation}
provided the "coefficients" $\alpha_i$ and $\beta$ obey the following algebra:
\begin{equation}
\alpha_i^2=\beta^2=1  , \hspace{10mm}  \{ \alpha_i,\beta \}=0, \hspace{10mm}  \{ \alpha_i,\alpha_j \}= 2 \delta_{ij} .
\end{equation}
The Pauli matrices
\begin{equation}
\sigma_1 =  \begin{pmatrix}
  0 & 1 \\
  1 & 0
 \end{pmatrix}
, \hspace{8mm}  \sigma_2 =  \begin{pmatrix}
  0 & -i \\
  i & 0
 \end{pmatrix} , \hspace{8mm}
 \sigma_3 =\begin{pmatrix}
  1 & 0 \\
  0 & -1
 \end{pmatrix},  
\end{equation}
which were introduced originally in order to describe the spin of the electron, satisfy this algebra. Nevertheless there are only $(2^2-1)/2=3$ Pauli matrices, whereas 4 anticommuting matrices are needed in 3+1 space-time: one for each space dimension and one for the mass. Hence Pauli matrices are not sufficient and it is necessary to use a higher representation consisting in four by four matrices. This is provided for instance by the so-called ordinary (or standard) representation defined by the following matrices:
\begin{equation}
\alpha_i =  \sigma_i \tau_1  =  \begin{pmatrix}
  0 & \sigma_i \\
  \sigma_i & 0
 \end{pmatrix},  \hspace{10mm}  \beta = \sigma_0 \tau_3 =\begin{pmatrix}
  \sigma_0 & 0 \\
  0 & -\sigma_0
 \end{pmatrix},
\end{equation}
which are here expressed in terms of tensor products between Pauli matrices $\sigma_i$ and $\tau_i$. Alternatively the Dirac equation can be written in the covariant form:
\begin{equation}
(i \gamma^\mu \partial_\mu - m ) \Psi= 0 ,
\end{equation}
using the Dirac matrices $\gamma^\mu=(\beta,\beta \alpha_i)$ and units where $\hbar=c=1$. The gamma matrices satisfy the Clifford algebra:
\begin{equation}
\{\gamma^\mu,\gamma^\nu \}=2 \eta^{\mu \nu},
\end{equation} 
where $\eta^{\mu \nu}$ is the Minkowski metric tensor ($\eta^{00}=-\eta^{ii}=1$). Therefore the gamma matrices in the ordinary representation read:
\begin{equation}
\gamma^i =  i \sigma_i  \tau_2  =  \begin{pmatrix}
  0 & \sigma_i \\
  -\sigma_i & 0
 \end{pmatrix},  \hspace{10mm}  \gamma^0  =\begin{pmatrix}
  \sigma_0 & 0 \\
  0 & -\sigma_0
 \end{pmatrix}.
\end{equation}

\subsection{Charge conjugation, space inversion and time-reversal}

The Dirac equation for a free electron with charge $q=-e$ in an external electromagnetic field (given by the potential vector $A_i$) can be written:
\begin{equation}
i\hbar \frac{\partial \Psi(\vr,t)}{\partial t} = \left[  \alpha_i (-i \partial_i - q A_i) + \beta m \right] \Psi(\vr,t) ,
\label{ED}
\end{equation}
in units where $\hbar=c=1$. Taking the complex conjugate of the above equation, one obtains: 
\begin{equation}
i\hbar \frac{\partial \Psi^*(\vr,t)}{\partial t} = \left[  \alpha_i^* (- i \partial_i + q A_i) - \beta^* m \right] \Psi^*(\vr,t) .
\label{ED2}
\end{equation}
Hence the wave function $ \Psi^*(\vr,t)$ obeys {\it almost the same original Dirac equation} as $\Psi(\vr,t)$, but with the following major difference: the charge $q$ in Eq. (\ref{ED}) is now replaced by the charge $(-q)$ in Eq. (\ref{ED2}). An unitary transformation $U_C$ is further needed for $U_C \Psi^*(\vr,t)$ to satisfy {\it exactly} the same Dirac equation as Eq. (\ref{ED}) with the simple replacement $q \rightarrow -q$. The requirements on $U_C$ read:
\begin{equation}
(U_C)^{-1} \alpha_i^* U_C = \alpha_i,  \hspace{10mm}  (U_C)^{-1}  \beta^* U_C = -\beta .
\end{equation}
Hence in the standard representation $U_C$ commutes with $\alpha_1$ and $\alpha_3$, but anticommutes with $\alpha_2$ and $\beta$. The only possibility is $U_C = i \tau_2 \sigma_2 =\gamma^2 $, so the full charge conjugation operator is given by the antiunitary transformation $\gamma^2 K_c$, where $K_c$ is the complex conjugation \cite{Zee:2010,Weinberg}.

The charge conjugation operation relates the wave function of the particle (charge $q$) to the wave function of its antiparticle (charge $-q$). The other discrete symmetries are space inversion $P$ and time-reversal $T$. Following a similar procedure as above, one finds that $P$ and $T$ are respectively associated to the unitary transformation $U_P=\gamma^0$ and $U_T=\gamma^1 \gamma^3 K_c$ \cite{Zee:2010,Weinberg}. Finally the product $PCT$ is always a symmetry and it is associated to the matrix $\gamma^5=i\gamma^0 \gamma^1 \gamma^2 \gamma^3$. In the ordinary (or Dirac) representation, this chirality or "handedness" matrix reads:
\begin{equation}
\gamma^5 =\sigma_0 \tau_1 = \begin{pmatrix}
  0 & \sigma_0 \\
 \sigma_0 & 0
 \end{pmatrix}.
\end{equation}
Finally, one can check easily that the charge conjugation and time-reversal transformations "square differently": $(U_C K_c)^2=1$ and $(U_T K_c)^2=-1$. We have used $(\gamma^i)^2=-1$ ($i=1,2,3$), and the fact that $\gamma^2$ is purely imaginary while $\gamma^1$ and $\gamma^3$ are real matrices. 

\subsection{Weyl equation for massless particles: helicity and chirality}

We now introduce another representation, the so-called chiral (or spinor) representation, which is very useful for fast (ultrarelativistic) particles, and in particular for massless particles as we will see below. This chiral representation is defined by the following matrices:
\begin{equation}
\alpha_i =  \sigma_i \tau_1  =  \begin{pmatrix}
  \sigma_i & 0 \\
  0 & -\sigma_i
 \end{pmatrix},  \hspace{10mm}  \beta = \sigma_0 \tau_1 =\begin{pmatrix}
 0 & \sigma_0 \\
  \sigma_0 & 0
 \end{pmatrix},
\end{equation} 
while the corresponding gamma Dirac matrices are:
\begin{equation}
\gamma^i=\beta \alpha_i =  \sigma_0 \tau_3 \sigma_i \tau_1  =\sigma_i (i\tau_2) = \begin{pmatrix}
  0 &  \sigma_i \\
  - \sigma_i & 0
 \end{pmatrix},  \hspace{10mm}  \gamma_0 = \beta = \sigma_0 \tau_1 , \hspace{10mm}  \gamma^5 = -\sigma_0 \tau_3 .
\end{equation} 
In this chiral representation, the Dirac equation for $\Psi=(\Psi_L,\Psi_R)$ reads:
\begin{align}
(i \sigma_0 \partial_t + i c \sigma_i \partial_i) \Psi_R &= (mc^2/\hbar) \Psi_L,\\
(i \sigma_0 \partial_t - i c \sigma_i \partial_i) \Psi_L &= (mc^2/\hbar) \Psi_R ,
\label{Weylcoupled}
\end{align}
and therefore the mass appears as a coupling between the 2-component spinors $\Psi_R$ and $\Psi_L$. The typical length $\hbar/(mc)$ that shows up naturally in these coupled equations is the Compton length. We shall see counterparts of this Compton length in condensed matter systems when studying the edge states confined between two insulators with typical energy gap $E_g$ and whose spatial extension will be roughly $\hbar v_F/E_g$ (Sec. \ref{SectionEdge}).   

\medskip

For a massless particle ($m=0$), the two-component Pauli spinors $\Psi_R$ (and $\Psi_L$) become decoupled and the equations become scale-invariant. Only three $\alpha_i$ matrices are needed in the Hamiltonian $H_D$ corresponding to the three directions of space (the $\beta$ matrix is not needed because $m=0$). Hermann Weyl was the first to notice that these equations correspond to two decoupled representation of the Lorentz group \cite{Zee:2010,Weinberg}. From Eq. (\ref{Weylcoupled}), one gets
\begin{equation}
( \sigma_0 \partial_t  -   c \boldsymbol{\sigma} . \partial_\vr )\Psi_L = 0 , 
\label{WeylequationL}
\end{equation}
whose solution, $\Psi_L \simeq e^{i(\vp.\vr-Et)/\hbar}$, has the energy $E= - c \boldsymbol{\sigma} . \vp$. The positive energy solution has a spin projection which points opposite to the momentum $\vp$ ({\it i.e.} $\boldsymbol{\sigma} . \vp < 0$), hence the name "left-handed fermion" and subscript "$L$". Similarly the equation for a right-handed particle reads
\begin{equation}
 (\sigma_0 \partial_t  +  c \boldsymbol{\sigma} .  \partial_\vr )\Psi_R= 0,
\label{WeylequationR}
\end{equation}
leading to a dispersion relation $E=c \boldsymbol{\sigma} . \vp$ and a spin pointing in the direction of motion ({\it i.e.} $\boldsymbol{\sigma} . \vp > 0$) for the positive energy solutions. Hence a massless particle can be described by a two (complex) component spinor with definite helicity or chirality (identical notions for massless particle). The chirality is the sign of the spin projection along the direction of the momentum of the particle. The chirality is a well-defined (i.e. a frame independent concept) only for massless particles. Indeed for a massive particle it is always possible to reverse the momentum $\vp$ (by choosing an inertial frame moving faster than the particle in the initial frame) while leaving the spin unchanged.

For several decades, it was unclear whether the neutrino was massless or has a tiny mass. Neutrino oscillation experiments seem to indicate a finite mass for the neutrino, which can be seen as coupling between the two Weyl equations above. Nowadays, the ongoing debate is on the structure of this coupling: Dirac type or Majorana type \cite{Avignone:2008}. For a Dirac mass, $\Psi_L$ and $\Psi_R$ are independent 2-component spinors. For a Majorana mass, $\Psi_L$ and  $\Psi_R=U_C \Psi_R$ are tied together by the charge conjugation symmetry $U_C$  \cite{Zee:2010,Weinberg,Pal:2010}.

\section{Graphene  \label{sectionGraphene}}
\label{SectionGraphene}

In high-energy physics, the relativistic wave equations described in previous section have to be understood in the framework of quantum field theory, in order to account for inelastic processes such as particle-antiparticle pair creation. In the context of condensed matter systems, the Dirac and the Weyl equations can be used directly in the first quantization formalism in order to describe the unique band structure of Dirac materials at low energy. We now turn to realizations of the Weyl equation for massless fermions in a famous two-dimensional material: graphene. Note that three dimensional realizations are provided by the Weyl semimetals, described in another contribution of this topical issue (Pavan Hosur and Xiaoliang Qi, {\it Recent developments in transport phenomena in Weyl semimetals}).

Graphene, the atomic-thin layer of carbon atoms, was first isolated on an insulating substrate in 2004 by two groups, in Manchester  University \cite{Novoselov:2004,Novoselov:2005} and Columbia \cite{Zhang:2005}, respectively. Before those milestone experiments, it was already predicted that graphene should host massless Dirac-Weyl fermions \cite{Wallace:1947,DiVincenzo:1984,Semenoff:1984,Haldane:1988} realizing the 2D Weyl equations Eq. (\ref{WeylequationL},\ref{WeylequationR}) where the velocity of light is replaced by a Fermi velocity $v_F \simeq c/300$. This relativistic character is remarkable considering that spin-orbit coupling is very weak in graphene. In fact the relativistic-like behavior originates from the particular honeycomb lattice structure that gives rise to a "momentum-isopin" coupling ("$\vp.\boldsymbol{\sigma}$" term in the Hamiltonian). As a consequence, the isospin $\boldsymbol{\sigma}$ involved in the relativistic dynamics is the sublattice index, and not the real electronic spin (which stays decoupled from orbital motion). Those experiments evidenced the existence of 2D Dirac-Weyl fermions by the measurement of a very particular Quantum Hall effect \cite{Novoselov:2005,Zhang:2005}, which is specific to relativistic carriers. It was also demonstrated that the density of such carriers can be tuned using a remote electrostatic gate, thereby realizing the first graphene-based field effect transistors \cite{Novoselov:2004}. Field effect transistors have a strong potential for applications in electronic devices, but they are also ideal systems to investigate the scattering properties of Dirac particles, including Klein tunneling \cite{Klein:1929,Cheianov:2006,Katsnelson:2006,Cayssol:2009,Yamakage:2011} which has been observed experimentally \cite{Huard:2007,Williams:2007,Ozylmaz:2007,Huard:2009,Young:2009}. The absence of backscattering at normal incidence is a property related to the symmetry of the spinor wave functions \cite{Ando:1998}. This protection against backscattering is rather "weak" in the sense that it assumes the absence of intervalley scattering, see Ref. \cite{Alain:2011}.

The physics of Dirac fermions in graphene has been extensively reviewed in far more details elsewhere \cite{CastroRMP:2009,GoerbigRMP:2011,KotovRMP:2012}. Here we recall how massless fermions appear as low-energy excitations of graphene, and set-up notations for the following sections. We discuss the protection of the Dirac points by the fundamental symmetries of the material.

\subsection{Tight-binding model}

\medskip

Graphene consists of a honeycomb lattice of carbon atoms with two interpenetrating triangular sublattices, respectively denoted A and B (Fig. 1). In this structure, each carbon atom has six electrons: two electrons filling the inner shell $1s$, three electrons engaged in the 3 in-plane covalent bonds in the $sp^2$ configuration, and a single electron occupying the $p_z$ orbital perpendicular to the plane. Much of the physics of graphene is related to the (2D) two-dimensional fluid formed by those $p_z$ electrons.   
\begin{figure}
\begin{center}
\includegraphics[width=6cm]{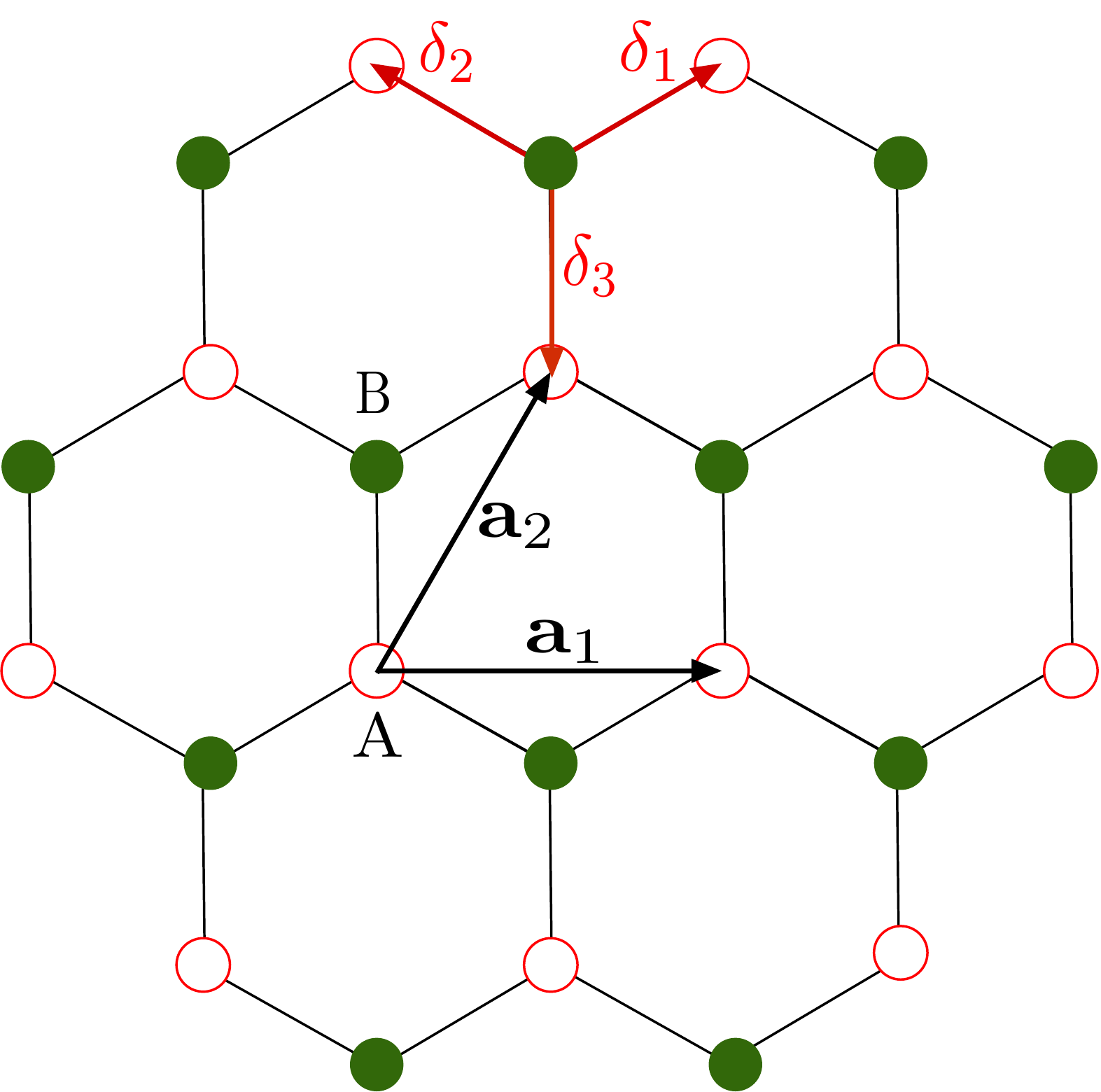}
\hspace{15mm}
\includegraphics[width=8cm]{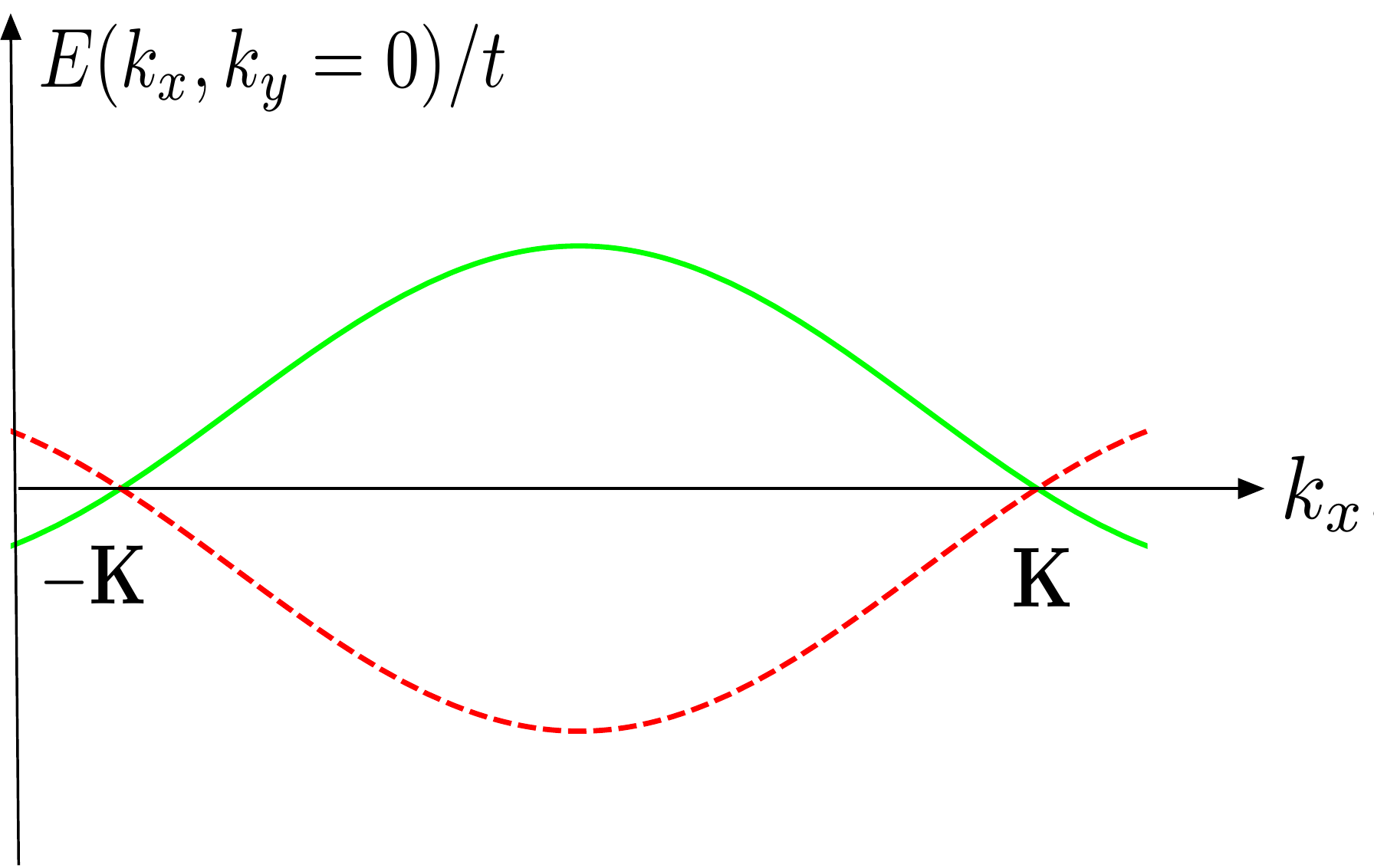}

\caption{{\it Left panel}: Graphene honeycomb lattice structure. Red open (green filled) dots for A (B) sublattice. The red thick arrows denote the vectors $\boldsymbol{\delta}_\alpha$ ($\alpha=1,2,3$) connecting of a given site to its three nearest neighbors. The black arrows are the basis vectors $\va_{1}$ and $\va_{2}$ of the Bravais lattice. The distance between two sites is $a=0.142$ nm and the surface of the unit cell is $A_{cell}=3 \sqrt{3} a^2 /2$. {\it Right panel}: Section of the electronic energy dispersion $E(\vk)=\pm |\vd(\vk)|$ of graphene for $k_y=0$, showing the two Dirac points at $\vk=\pm \vK$.}
\end{center}
\label{FigHoneycomb}
\end{figure}
It is thus natural to use a single orbital tight-binding Hamiltonian:
\begin{equation}
\label{GrapheneHamiltonian}
H_0  =  t \sum_{\vr_A} \sum_{\alpha=1}^{3}   c_B^\dagger(\vr_A+\boldsymbol{\delta}_\alpha) c_A(\vr_A) + {\rm H.c.},
\end{equation}
where $t \simeq -2.7$ eV is the hopping amplitude between the $p_z$ orbitals of two adjacent carbon atoms. The operator $c_a(\vr_i)$ destroys a fermion in the orbital $p_z$ at site $\vr_i$, with $a=A,B$ indicating the sublattice. The sum over $\vr_A$ runs over the A-sites which form a triangular Bravais lattice spanned by the basis vectors:
\begin{equation}
\va_{1}=   \sqrt{3}  a \, \vec{e}_x ,     \,   \va_{2}=  \frac{a}{2} \left(    \sqrt{3} \vec{e}_x + 3 \vec{e}_y \right)    ,
\end{equation}
where $a=0.142$ {\rm nm} is the length of the carbon-carbon bond. The vectors $\boldsymbol{\delta}_{\alpha}$ defined by 
\begin{equation}
\boldsymbol{\delta}_{1,2}=   \frac{a}{2} \left(   \pm \sqrt{3} \vec{e}_x +  \vec{e}_y \right)   ,     \,   \boldsymbol{\delta}_{3}=   -  a \, \vec{e}_y ,
\end{equation}
connect any A-site to its three B-type nearest neighbors (Fig. \ref{FigHoneycomb}). The hopping matrix elements between next-nearest neighbors are neglected, which is justified by the fact that those corrections are roughly ten times smaller than the main hopping $t$. 
  
Owing to translation invariance, the two-dimensional quasi-momentum $\vk =(k_x,k_y)$ is a good quantum number. In order to diagonalize the Hamiltonian Eq. (\ref{GrapheneHamiltonian}), we use the Fourier transformation:  
\begin{equation}
\label{Fourier}
c_a(\vr_i)= \frac{1}{\sqrt{N}} \sum_{\vk} e^{-i \vk . \vr_i} c_{a}(\vk), 
\end{equation}
where $a=A,B$ is the sublattice index and $N$ is the total number of sites. After substitution of Eq.(\ref{Fourier}), the Hamiltonian Eq. (\ref{GrapheneHamiltonian}) becomes diagonal in momentum and reads: 
\begin{equation}
\label{AGrapheneHamiltonianFourier}
H_0  = \sum_\vk    c_a^\dagger(\vk) [h_0(\vk)]_{ab} \, c_b(\vk),
\end{equation}
where $\vk$ is restricted to the first Brillouin zone (BZ). The Bloch Hamiltonian $h_0(\vk)$, which acts on the sublattice isospin, is given by 
\begin{equation}
h_0(\vk)=d_1(\vk) \sigma_1 + d_2(\vk) \sigma_2 ,
\label{hzerographene}
\end{equation}
since only off-diagonal hopping amplitudes are included in the model defined by Eq. (\ref{GrapheneHamiltonian}). The real functions $d_1(\vk)$ and $d_2(\vk)$ are defined by:
\begin{equation}
d_1 (\vk)  + i d_2 (\vk)   =  t \sum_{\alpha=1}^{3} \exp(i \vk . \boldsymbol{\delta}_\alpha) ,
\label{EqEvenOdd}
\end{equation}
over the whole BZ. The functions $d_1 (\vk)$ and $d_2 (\vk)$ are respectively even and odd in momentum which will be important for the symmetry analysis in Sec. \ref{symm}. 
The electronic energy spectrum is given by the length of the vector $\vd =(d_1,d_2)$:
\begin{equation}
\label{SpectrumTBM}
E(\vk)=\epsilon_0(\vk) \pm |\vd(\vk)| = \pm \sqrt{d_1^2 (\vk)+d_2^2 (\vk)},
\end{equation}
which describes a valence band (minus sign) and a conduction band (plus sign) that are symmetric with respect to $E=0$. The zero energy corresponds to the common energy of the $p_z$ atomic orbitals on sublattices $A$ and $B$. The valence and conduction bands touch at isolated points of the Brillouin zone obtained by solving  the equation $\vd (\vk)=0$. There are only two inequivalent Dirac points located at:
\begin{equation}
\vk = \pm \vK=\pm \frac{4 \pi}{3\sqrt{3}a} \, \vec{e}_x ,
\end{equation}
in reciprocal space. Other solutions of the equation $\vd(\vk)=0$ can be linked by a reciprocal lattice vector to one of these two solutions, and therefore describe the same physical state. 

\medskip

The existence of isolated solutions of  $\vd(\vk)=0$, preventing the system to become gapped, is robust even if some crystal symmetries are lost and more hopping amplitudes are added. For instance additional second-neighbor hoppings will break the electron/hole symmetry discussed above, but will not affect the existence of Dirac points. Other perturbations, like a anisotropic deformations on one type of bond, only shift the Dirac points and modify the conical dispersion around them \cite{Goerbig:2008,guinea:2010,Ghaemi:2012}. In fact the touching points are protected by more fundamental symmetries, namely space inversion and time-reversal symmetries.

\subsection{Low energy effective theory near the Dirac points}
\label{lowenergyH}

We consider now the low-energy theory for the single-particle states near the Dirac points. The momenta are written as $\vk =\pm  \vK + \vq$ close to the zero-energy points ($|\vq|a \ll 1$), and the annihilation operators for these states are denoted $c_{A\pm\vK} (\vq)=c_{A}(\pm\vK+\vq)$, where $\vq = q_x \ve_x +  q_y \ve_y$ is a small momentum deviation from the Dirac points. Expanding to first order in momenta, the Hamiltonian describing the low energy excitations near $\vk=\xi \vK$ ($\xi=\pm 1$) is found to be:
\begin{equation}
H_0^{(\xi \vK)} =v_F \sum_\vk
\begin{pmatrix} 
c_{A\xi\vK}^\dagger(\vq) & c_{B\xi\vK}^\dagger (\vq)   \\
\end{pmatrix}
\begin{pmatrix} 
0 & \xi q_x -i q_y \\
 \xi q_x  + i q_y & 0 
\end{pmatrix}
\begin{pmatrix} 
c_{A\xi\vK} ( \vq) \\
c_{B\xi\vK} (\vq) 
\end{pmatrix},
\end{equation}
where $v_F = -3 at/2 \simeq 10^6$ m.s$^{-1} \simeq c/300$ is the Fermi velocity. The Fermi velocity is basically the bandwidth $t$ divided by the Brillouin zone (BZ) size $1/a$. Therefore near each of the Dirac points, one obtains a 2D Weyl Hamiltonian describing massless relativistic particles. Using the convenient spinor representation, $
c_\alpha^\dagger(\vq)=( c_{A\vK}^\dagger  c_{B\vK}^\dagger   c_{A-\vK}^\dagger  c_{B-\vK}^\dagger  ) $,
the single-electron Hamiltonian can be written in the compact form:
\begin{equation}
H_0 = \sum_{\vq} \sum_{\alpha,\beta=1}^{4} c_\alpha^\dagger(\vq) [v_F  (q_x  \sigma_1 \tau_3+ q_y  \sigma_2)]_{\alpha \beta} \, c_\beta(\vq) ,
\label{Hamiltonianzerospinor}
\end{equation}
which has exactly the form of the Dirac Hamiltonian describing a spin one-half relativistic particles with zero mass. In particular the dispersion relation is simply:
\begin{equation}
E(\vq) =v_F  |\vq| ,
\label{DispersionMasslessDirac}
\end{equation}
typical of a relativistic massless particle with velocity of light replaced by $v_F$.

Nevertheless we would like to emphasize again the differences between the Dirac equation in the contexts of graphene and particle physics, respectively. In high-energy physics, the Dirac equation comes from Lorentz-invariance and very general considerations related to special relativity and quantum mechanics (Sec. \ref{SectionWave}). Then, in 3+1 space-time dimensions, the minimal objects to satisfy Dirac equation are pairs of bispinors combining the spin and particle/antiparticle degrees of freedom.

In graphene, the origin of the Dirac physics is totally different. As we have seen, the spinors originate from a $\vk .\vp$ expansion around special points of a particular band structure. Hence in graphene, there is no fundamental issue with the negative energy states that are just the valence band states (these states are in fact bounded from below by the bottom of the valence band). Finally the emergent Lorentz invariance of Eq. (\ref{DispersionMasslessDirac}) is only valid near the Dirac point, namely for wave vectors $\vq$ located in a disk whose radius is far smaller than the inverse lattice spacing $1/a$, whereas Lorentz invariance applies in the whole Minkowski space-time in particle physics. Finally the 4 components of the spinors are associated to the sublattice isospin (instead of real spin), and to the valley index (instead of particle/antiparticle label).


Fundamentally graphene is a two-band system because it has 2 orbitals per unit cell (one $p_z$ orbital per atom and 2 atoms in the unit cell), thereby having two states per momentum $\vk$ in the first Brillouin zone (BZ). Two species of massless Dirac fermions (one for each valley) which carry a sublattice isospin coupled to their momentum. This is an illustration of fermion doubling on a lattice. Note that the presence of 4$\times$4 matrices in Eq.(\ref{Hamiltonianzerospinor}) does not mean that graphene is a 4 band system in the same sense as the genuine 4-band insulators. Indeed for a given value of $\vk$ in the BZ graphene has only two states (Fig. 1, right panel). Besides, when discussing transport or at least ballistic elastic scattering at the Fermi level, there are effectively 4 states sharing the same energy (Fig. 1, right panel). Then the physics depends on the ratio between intravalley and intervalley scattering rates. For instance, Klein tunneling and weak antilocalization are better observed if the intervalley coupling is much weaker than the intravalley coupling.  

Here we have not considered explicitly the real spin simply because it is not coupled to the momentum in the absence of spin-orbit. In fact at each valley, there are two completely degenerated and decoupled Dirac cones corresponding to each spin direction. Hence graphene has 4 Dirac cones with sublattice-momentum locking. This is at odds with surface state of 3D strong topological insulators which has a single Dirac cone with momentum coupled to the real spin.

\subsection{Symmetries: space inversion and time reversal} 
\label{symm}

Here we discuss the robustness of the Dirac points in graphene. It turns out that those Dirac points are remarkably robust as long as some fundamental symmetries are obeyed and spin-orbit is weak \cite{Bernevig:2013}. These fundamental symmetries are time-reversal symmetry $\mathcal T$ and inversion symmetry $\mathcal P$. Again we restrict our discussion to spinless fermions.   

The inversion symmetry $\mathcal P$ switches the sublattice $A$ and $B$, and therefore should transform the Pauli matrices as:
\begin{equation}
\mathcal P:  \hspace{4mm}  (\sigma_1,\sigma_2,\sigma_3) \rightarrow (\sigma_1,-\sigma_2,-\sigma_3),
\end{equation}
while the time-reversal operation leaves invariant the sublattice but complex conjugates the wave functions amplitudes, acting therefore as:
\begin{equation}
\mathcal T:  \hspace{4mm}  (\sigma_1,\sigma_2,\sigma_3) \rightarrow (\sigma_1,-\sigma_2,\sigma_3),
\end{equation}
Hence the inversion and time-reversal operations will be written as:
\begin{equation}
\mathcal P =\sigma_1  ,  \hspace{10mm}   \mathcal T =\sigma_0 K_c ,
\end{equation}
where $K_c$ is complex conjugation. It is important to note that this time-reversal operation obeys $\mathcal{T}^2=1$ because we deal with spinless fermions. When the spin is included, the full time-reversal operation square to $-1$ with Kramers degeneracy as a fundamental consequence. It is clear that the form of the symmetry operations $\mathcal T$ and $\mathcal P$ depends on the fact that the Pauli matrices $\sigma_i$ represent the sublattice isospin \cite{Bernevig:2013}, and not real electronic spin. In particular the expressions for $\mathcal T = \sigma_0 K_c$ and $\mathcal P=\sigma_1$  differ from the ones obtained in particle physics ($U_T$ and $U_P$ in Sec. 2.2).  

In order to discuss the stability of the Dirac points, we might investigate how a perturbation like $d_3(\vk) \sigma_3$ transforms under application of $\mathcal P$ and $\mathcal T$. Invariance under $\mathcal P$, implies $d_3(\vk)=-d_3(-\vk)$ while invariance under $\mathcal T$, implies $d_3(\vk)=-d_3(-\vk)$. This means that such a perturbation breaks either inversion or time-reversal symmetry, and we would study in detail the corresponding graphene insulators below (Sec. \ref{insulators}).    

Now we can check that the Hamiltonian is invariant both under the time-reversal operation $\mathcal T$ and inversion $\mathcal{P}$:
\begin{equation}
\mathcal{T} h_{0}(\vk) \mathcal{T}^{-1}= \left(   d^*_1(\vk)  \sigma_1 - d^*_2(\vk)  \sigma_2  \right)=\left(   d_1(-\vk)  \sigma_1 + d_2(-\vk)  \sigma_2  \right)= h_{0}(-\vk),
\end{equation}
\begin{equation}
\mathcal{P} h_{0}(\vk) \mathcal{P}^{-1}=  \sigma_1  \left(  d_1(\vk) \sigma_1 + d_2(\vk) \sigma_2  \right) \sigma_1   = h_{0}(-\vk) ,
\end{equation}
because the real functions $d_1(\vk)$ and $d_2(\vk)$ are respectively even and odd in momentum as shown by Eq. (\ref{EqEvenOdd}).

\medskip

Finally in the low energy effective theory, representations gathering the valley and the sublattice isospins are often used. Then one has to take into account that space inversion and time-reversal operations also switch the valleys. For instance, in the representation $c_{a}^\dagger(\vq)=( c_{A\vK}^\dagger  c_{B\vK}^\dagger   c_{A-\vK}^\dagger  c_{B-\vK}^\dagger  )$ introduced in \ref{lowenergyH}, the space inversion and time-reversal operations are four by four matrices: 
\begin{equation}
\mathcal P =\sigma_1 \tau_1 , \hspace{1cm}    \mathcal T = \sigma_0 \tau_1 K_c ,
\end{equation}
involving a supplementary Pauli matrix $\tau_1$ for the valley inversion.

\section{Masses in graphene: Semenov, Haldane, Kane-Mele insulators}
\label{insulators}

Semimetallic graphene exhibits robust massless Weyl-Dirac fermions protected as long as space inversion and time-reversal symmetries are not broken. In this section we discuss how to generate gaps at the Dirac points and emphasize the concept of mass in graphene. Owing to the sublattice isospin, distinct insulating phases can be built (at least theoretically) by adding proper perturbations \cite{Ryu:2009}. 

\subsection{Two-band model}
A mass term is a matrix that acts on the sublattice isospin and anticommutes with the Hamiltonian of semimetallic graphene. When discussing the case of spinless fermions on a bipartite lattice, the only matrix anticommuting with the "velocity" matrices ($\sigma_1$ and $\sigma_2$ in Eq. (\ref{hzerographene})) is the third Pauli matrix $\sigma_3$. Due to the simplicity of the model, there is no choice on the matrix, but there are still many different functions $d_3(\vk)$ that can enter the Hamiltonian. Therefore the simplest and most generic model for spinless fermions on a bipartite lattice is the two-band model:
\begin{equation}
h(\vk)=\epsilon_0 (\vk) \sigma_0 + d_1(\vk) \sigma_1 + d_2(\vk) \sigma_2 +  d_3(\vk) \sigma_3 = \epsilon_0 (\vk) \sigma_0 +  \vd(\vk) .  \boldsymbol{\sigma}.
\label{2bandH}
\end{equation}
Of course materials usually have much more bands, but this model single out two bands where "interesting things" happen. By "interesting things" we mean at first vicinity of the Fermi level, and eventually a band-crossing that may induce non trivial topological properties  \cite{KaneRMP:2011,QiRMP:2011,Bernevig:2013}. This is in the same spirit as the two-level model in atomic physics. Nevertheless there is a crucial difference: in materials the coefficients of the Pauli matrices gain a $\vk$-dependence, which allows the possibility of Berry-Zak phase effects \cite{Zak:1989}. Moreover  including additional internal degrees of freedom usually leads to an increase in the number of various possible mass terms. For spinless fermions on the honeycomb lattice, there are only 4 possible mass terms (Semenov, Haldane and two Kekule distortions) which all break some symmetry. When the spin is included, there are 16 different masses, most of them breaking some symmetries while others, like the Kane-Mele mass, respect all the symmetries \cite{Ryu:2009}.

\subsection{Semenov insulator} 

The simplest choice consists in a constant mass term $d_3(\vk)=M_S$ (independent of $\vk$), which was first discussed by Semenov \cite{Semenoff:1984}. Such a mass corresponds physically to a staggered on-site potential that spoils the inversion symmetry (equivalence between A and B sites) while leaving time-reversal symmetry intact. It is realized for a honeycomb lattice where the A and B sites are actually occupied by distinct atoms, like BN crystals. In this review, we will call such gapped electronic system a Semenov insulator. This Semenov phase is also realized with cold atoms trapped in tunable optical lattices \cite{tarruell2011}. The spectrum of the bulk excitations is given by the relativistic dispersion relation $E^2=v_F^2 p^2 + M_S^2$, and there is no edge state. Hence this system is insulating both in the bulk and along its edges.

\subsection{Haldane model and Quantum Anomalous Hall insulators (spinless electrons)}
\label{HaldaneInsulatorSection}

\begin{figure}
\begin{center}
\includegraphics[width=6cm]{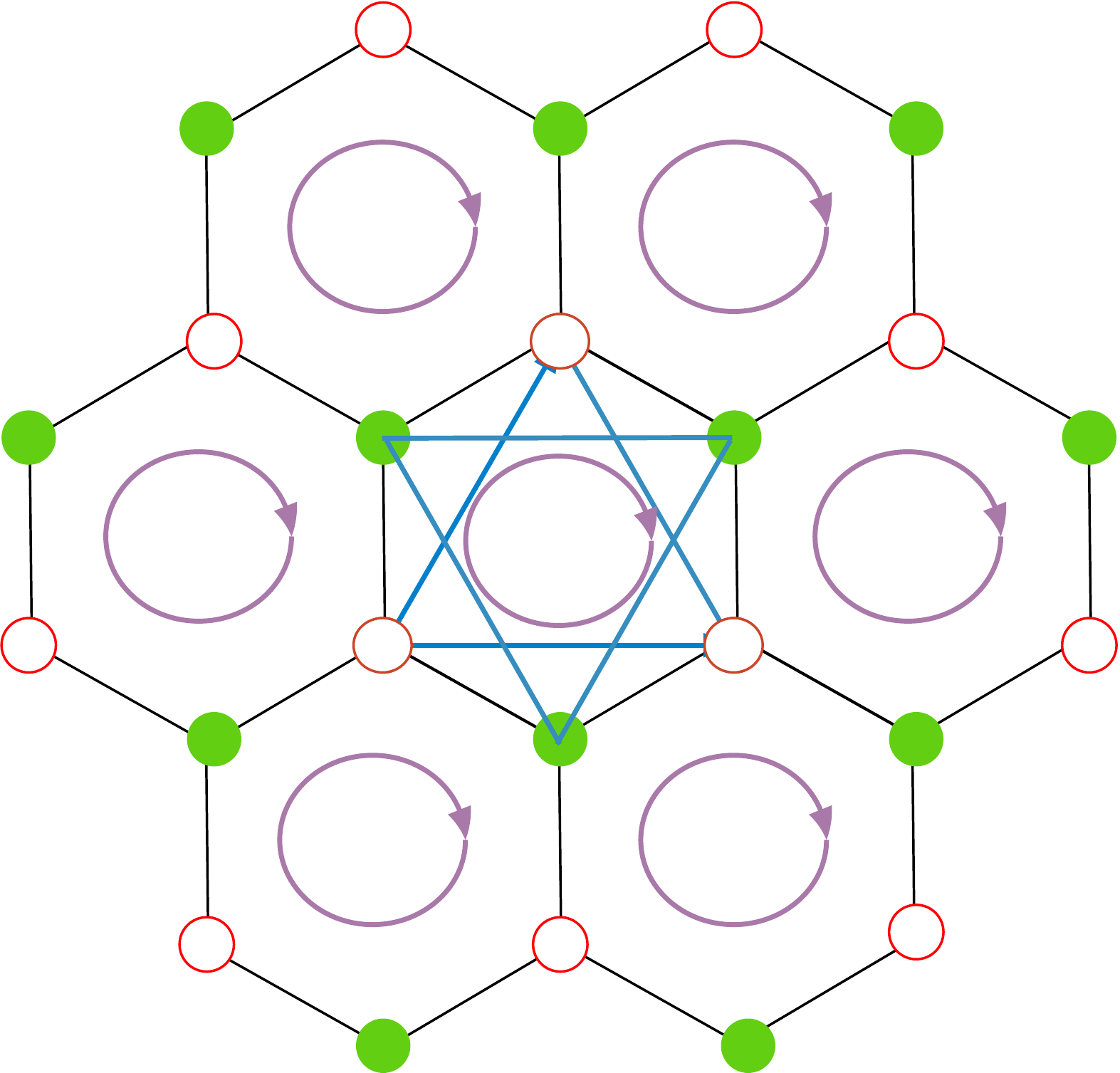}
\hspace{20mm} 
\includegraphics[width=6cm]{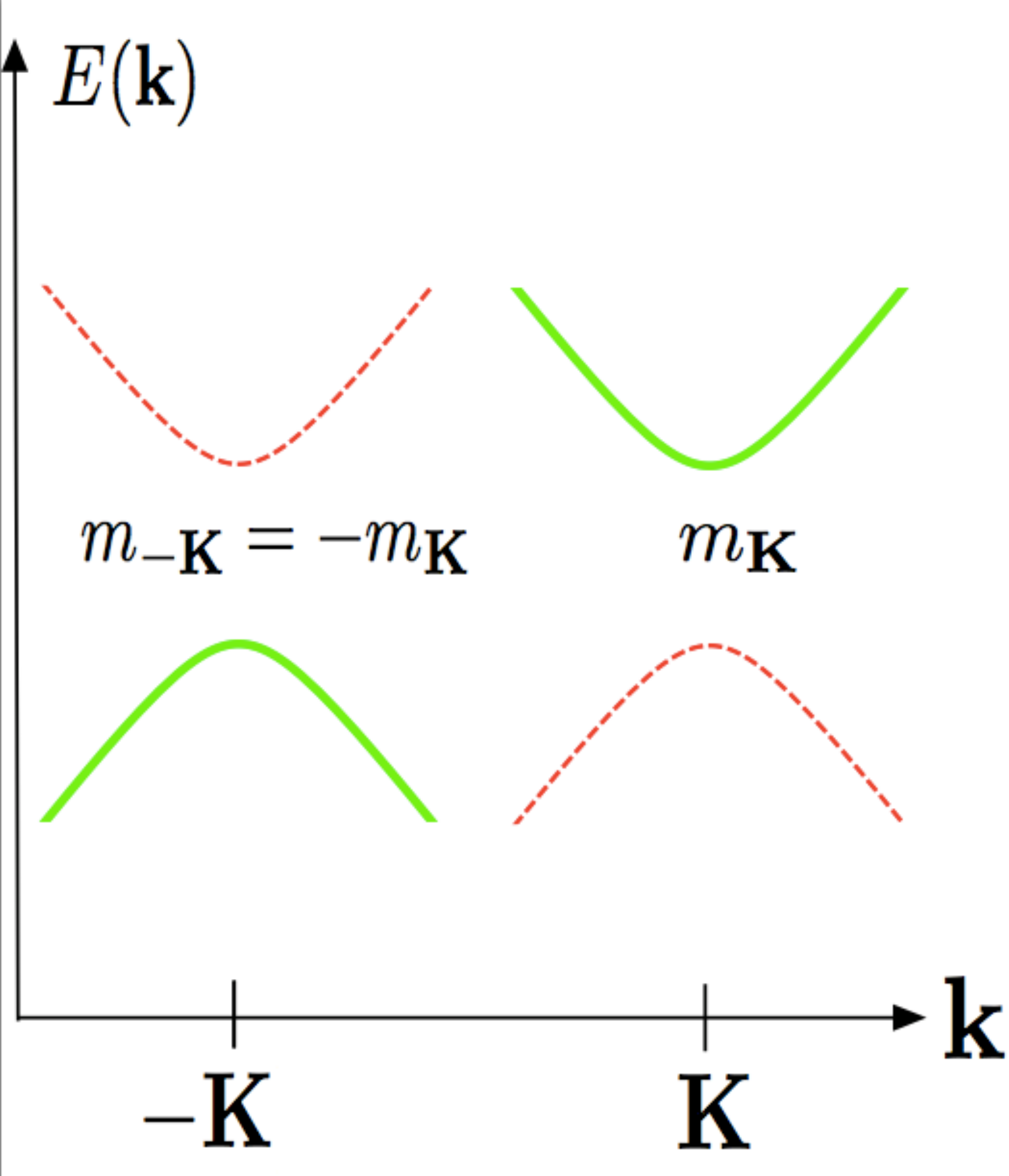}
\caption{{\it Left panel}: Next-nearest neighbor (NNN) couplings in the Haldane model of graphene. Blue arrows (direction indicated by the oriented loops inside the hexagons) stands for $t_2 e^{i \phi}$. The vectors connecting NNN neighbor sites are defined by $\vb_1=\boldsymbol{\delta}_{2}-\boldsymbol{\delta}_{3}$, $\vb_2=\boldsymbol{\delta}_{3}-\boldsymbol{\delta}_{1}$, and $\vb_3=\boldsymbol{\delta}_{1}-\boldsymbol{\delta}_{2}$. {\it Right panel}: The gaps at the Dirac points depend on the phase  $\phi$ ad $m_{\pm \vK}=d_3 (\vk \simeq \pm \vK)  = \mp 3 \sqrt{3} t_2 \sin (\phi) $.}
\end{center}
\label{FigHaldaneLattice}
\end{figure}
Motivated by the search of quantum Hall phases in the absence of the Landau level structure, Haldane proposed a Bloch band insulator model which exhibits quantized Hall conductance $\pm e^2/h$ \cite{Haldane:1988}.

\medskip

{\bf Model and bulk physics:} In order to realize a finite quantum Hall response with Bloch states, it is necessary to break time-reversal symmetry while preserving the translational symmetry of the Bravais lattice. This can be done by inserting local fluxes which sum up to zero over each unit cell. Such a pattern preserves the Bloch nature of electronic states. In the Haldane model of graphene, these fluxes can be described by introducing unimodular phase factors in the second neighbor hopping amplitudes $t_2\rightarrow t_2 e^{\pm i \phi}$, where the $\pm$ sign corresponds to the different chiralities (Fig. 2, left panel). To be more specific, the second-neighbor hoppings are: 
\begin{equation}
\label{GrapheneH2}
H_2 =  t_2   \sum_{i=1}^{3}   \left( \sum_{\vr_A} c_A^\dagger(\vr_A) c_A(\vr_A +\vb_i) e^{i \phi} +\sum_{\vr_B}   c_B^\dagger(\vr_B) c_B(\vr_B + \vb_i) e^{-i\phi} \right) + {\rm H.c.},
\end{equation}
where $\vb_1=\boldsymbol{\delta}_{2}-\boldsymbol{\delta}_{3}$, $\vb_2=\boldsymbol{\delta}_{3}-\boldsymbol{\delta}_{1}$, and $\vb_1=\boldsymbol{\delta}_{2}-\boldsymbol{\delta}_{3}$, are the vectors connecting next-nearest neighbor sites. After a Fourier transform, this Hamiltonian becomes 
\begin{equation}
\label{HaldaneHamiltonianFourier}
h_{2}(\vk)  =  \epsilon_0(\vk) \sigma_0  +  d_3 (\vk) \sigma_3   ,
\end{equation}
with:
\begin{equation}
\epsilon_0(\vk) =2 t_2 \cos (\phi)   \sum_{i=1}^{3}       \cos(\vk . \vb_i) ,  \hspace{10mm} d_3 (\vk)   =   2 t_2 \sin (\phi)   \sum_{i=1}^{3}    \sin(\vk . \vb_i)  .
\end{equation}
The NNN perturbation is dispersive ($\vk-$dependent) because it is nonlocal in real space. The part of $h_2(\vk)$ which is proportional to the identity just shifts the energies and spoils the electron-hole symmetry of the purely NN model. Nevertheless the system remains gapless (if $\phi=0,\pi$) under introduction of a real NNN hoppings, because both $\mathcal{P}$ and $\mathcal{T}$ are preserved for real NNN hoppings. In contrast, for complex hoppings, the term proportional to $\sigma_3$ opens gaps at the Dirac points. Near the Dirac points, one has simply to substitute $\vk=\vK$ (or $\vk=-\vK$) as a zero order approximation, and one obtains:
\begin{equation}
d_3 (\vk \simeq \pm \vK)  = \mp 3 \sqrt{3} t_2 \sin (\phi)   ,
\label{HaldaneTerm}
\end{equation}
using that:
\begin{equation}
\label{Sums}
\sum_{i=1,2,3}   \cos(\vK . \vb_i)   =  - \frac{3}{2} ,   \hspace{10mm}  \sum_{i=1,2,3}  \sin(\pm \vK . \vb_i) = \mp \frac{3 \sqrt{3}}{2} .
\end{equation}
This Haldane mass term $d_3(\vk)$ is odd in momentum and in particular changes sign in different valleys (Fig. 2, right panel). This is at odds with the Semenov insulator where both valleys are characterized by the same gap. The spectrum is again obtained by squaring the Bloch Hamiltonian, and using the properties of the Pauli matrices:
\begin{equation}
E^2  = v_F^2 p^2 + 27 t^2 \sin^2\phi .
\label{Spectrum}
\end{equation}
The gap/mass depends on the flux and cancels for $\phi=0$ where the time-reversal is trivially restored. This is also the case at $\phi=\pi$ because $e^{i\pi}=e^{-i\pi}=-1$.

\medskip

{\bf Chiral edge mode:} As we shall see in more details in Sec. \ref{SectionEdge}, the gapped bulk excitation coexist with a single edge state that carries the Hall conductance $\pm e^2/h$. There is no backscattering because the excitations propagate only in a single direction and there no states circulating in the opposite direction. 

\medskip

{\bf Quantum anomalous Hall (or Chern) insulators:} The Haldane model is representative of a wider class of 2D insulators exhibiting a finite Chern number and the QHE: the Chern insulators, also called the Quantum Anomalous Hall (QHA) insulators \cite{tang2011,sun2011,neupert2011,regnault2011,donna2011,neupert2}. Nevertheless such Chern insulators were so far difficult to realize experimentally since a nontrivial magnetic (or gauge field) texture is required. Recently, experimental evidence of the QHA state has been reported thin films of chromium-doped (Bi,Sb)$_2$Te$_3$ (Sec. \ref{experimental}) \cite{Chang:2013}. Note that those QHA (or Chern insulators) can also lead to incompressible states, similar to the fractional quantum Hall states, in presence of interactions and for partially filled flat bands. These so-called fractional Chern insulators are reviewed in in another contribution of this topical issue (S. A. Parameswaran, R. Roy, and S. L. Sondhi, {\it Fractional Quantum Hall Physics in Topological Flat Bands}).

\begin{figure}
\begin{center}
\includegraphics[width=6.5cm]{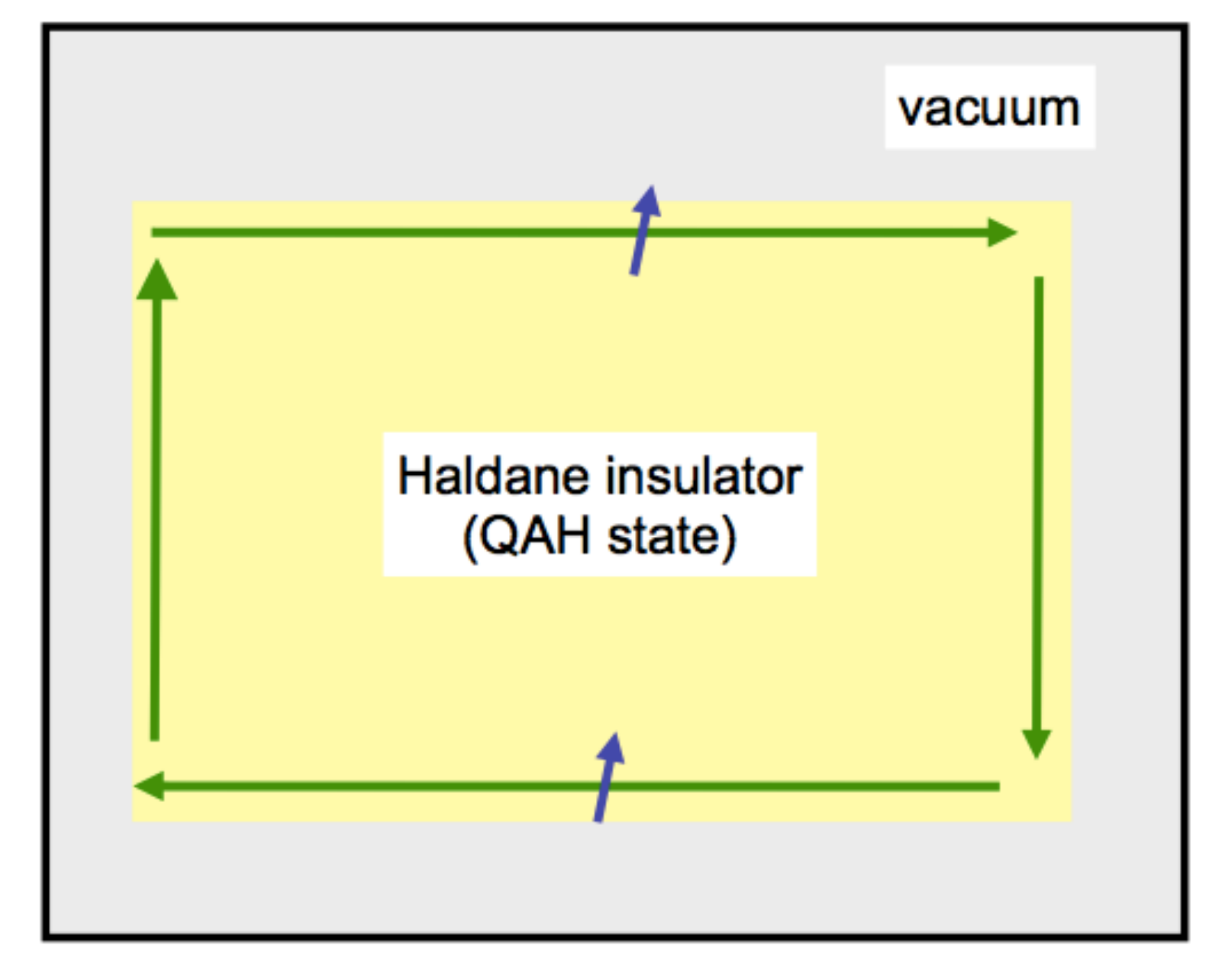}
\caption{Haldane insulator (see Sec. \ref{HaldaneInsulatorSection}) is characterized by a robust chiral edge state (green arrows). It belongs to the wider class of Chern insulators characterized by the presence of bands carrying a finite Chern number. The tilted (blue) arrows refer to the unique spin direction (fermions are assumed to be spinless, i.e. totally spin-polarised).}
\end{center}
\label{FigHaldaneInsulator}
\end{figure}

\subsection{Kane-Mele model and Quantum Spin Hall insulators (spinful electrons)} 
\label{KaneMeleInsulatorSection}
We have seen that the Haldane model for spinless fermions on the honeycomb lattice (as all other Chern insulator models) breaks time-reversal symmetry and exhibits the topologically protected integer quantum Hall effect \cite{Haldane:1988}. 
In 2005, C.L. Kane and E.G. Mele proposed a generalization of the Haldane model that respects time-reversal invariance and includes the spin via the spin-orbit interaction. Their idea launched the field of time-reversal invariant topological insulators which has known a rapid expansion since then \cite{KaneRMP:2011,QiRMP:2011,KonigJPSJ:2008,QiPhysToday:2010}. In the absence of spin-orbit coupling, the Dirac points of graphene are protected by the combination of two fundamental discrete symmetries: time-reversal and space inversion. This situation is drastically changed when the spin is coupled to electronic motion: intrinsic spin-orbit coupling does open a gap at the Dirac points without breaking any of those fundamental symmetries \cite{Kane:2005a,Kane:2005b}. The resulting insulator, the so-called Quantum Spin Hall (QSH), is a novel state of electronic matter that cannot be adiabatically connected to a trivial atomic insulator without closing (and re-opening) the bulk gap. Note that the terminology "Quantum Spin Hall insulator" can be misleading in the following sense: the QSH insulator refers to a new insulating state that is very different from the doped semiconductors exhibiting the so called intrinsic spin Hall effect \cite{Murakami:2003,Sinova:2004,Kato:2004,Wunderlich:2005}. Nevertheless, if the Fermi level is raised into the 2D conduction or valence bands, the Kane-Mele insulator becomes a doped semiconductor with strong Berry phase effects causing the intrinsic (band induced) spin Hall effect. 

\medskip

{\bf Bulk physics:} In their seminal paper, C.L. Kane and E.G. Mele first discussed the lattice model for spinful electrons on the honeycomb lattice  \cite{Kane:2005a} described by the Hamiltonian:  
\begin{equation}
\label{KaneMeleHamiltonian}
H=  t \sum_{\langle i,j \rangle}    c^\dagger_{i\alpha} c_{j\alpha} + i  t_2 \sum_{\langle \langle i,j \rangle \rangle}  \nu_{ij}   c^\dagger_{i\alpha} (s_{3})_{\alpha \beta} c_{j\beta},
\end{equation}
where $i,j$ denote the sites of the honeycomb lattice, the Pauli matrix $s_3$ refers to the physical spin of electrons, and the summation over repeated spin index ($\alpha,\beta$) is implied. The first term is a sum over the nearest-neighbor sites, denoted $\langle \langle i,j \rangle \rangle$, which defines the usual tight binding model for graphene (see Sec. \ref{SectionGraphene}). The second term, introduced by Kane and Mele, is a sum over next-nearest neighbors ($\langle \langle i,j \rangle \rangle$) where the hopping term $i t_2 \nu_{ij} s_3$ describes a spin-orbit coupling between the spin direction $s_3=\pm 1$ (units of $\hbar/2$) and the "chirality" $\nu_{ij}=\pm 1$ of the circulating electrons. This can be seen as a $L.S$ coupling where the "orbital momentum $L$" would be associated with the chirality. One can develop a very simple local picture for this spin-orbit coupling. 
\begin{figure}
\begin{center}
\includegraphics[width=6.5cm]{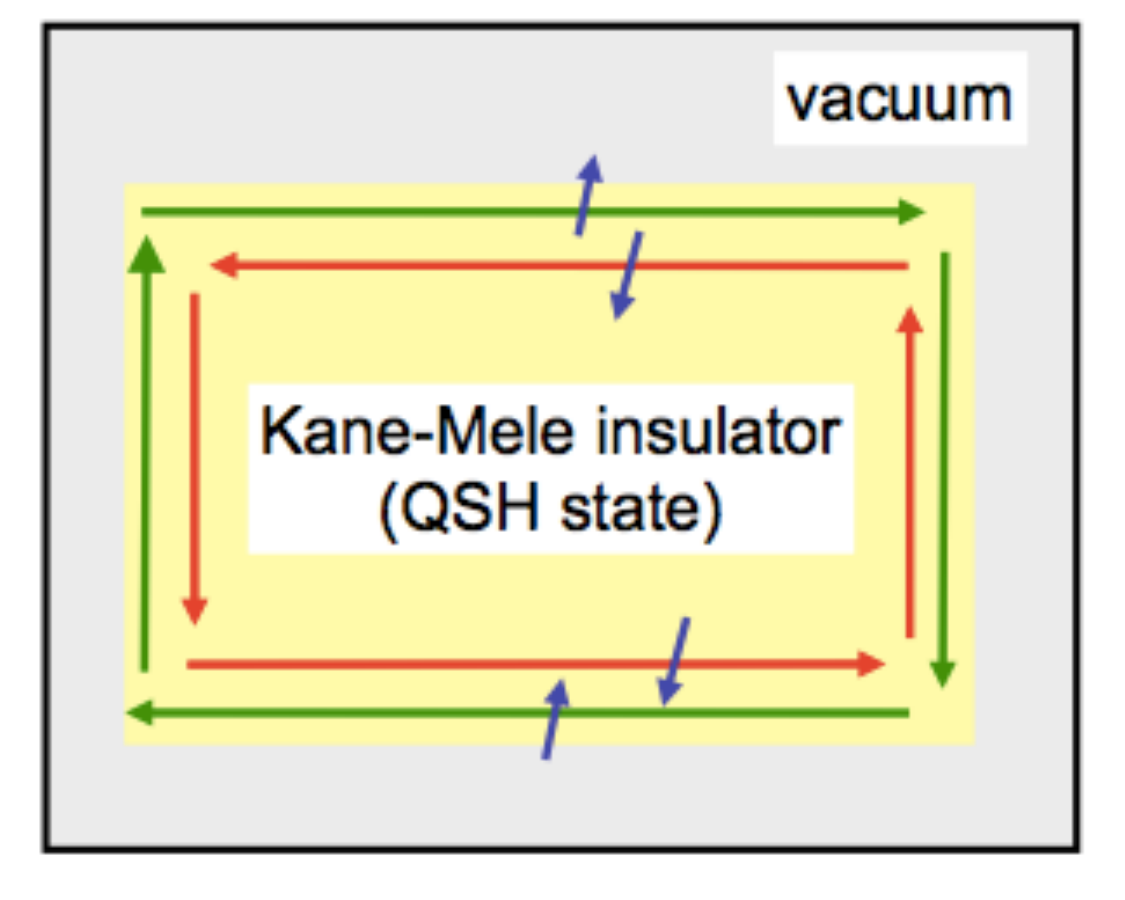}
\caption{The Kane-Mele insulator belongs to the wider class of Quantum Spin Hall insulators (Sec. \ref{KaneMeleInsulatorSection}) which are characterized by a Z$_2$ topological invariant in the bulk and a helical edge mode (represented schematically with the red and green arrows along the edge of the insulator). The tilted (blue) arrows refer to the two spin directions. There is a complete spin-momentum locking: spin-up electrons circulate clockwise while their Kramer partners circulate anticlockewise with spin-down. This helical mode is protected from backscattering (between clockwise and anti clockwise movers) as long as time-reversal symmetry is obeyed and as long as the bulk gap remains open.}
\end{center}
\label{FigKMInsulator}
\end{figure}
Since $[H,s_z]=0$, the model Eq.(\ref{KaneMeleHamiltonian}) can be decoupled in two subsystems for spin up ($s_3 =1$) and spin-down ($s_3=-1$) respectively. The Hamiltonian for spin-up (resp. spin-down) electrons is the Haldane Hamiltonian Eq.(\ref{GrapheneH2}) with $\phi=\pi/2$ (resp. $\phi=-\pi/2$). Hence many properties can be deduced from our knowledge of the Haldane model for spinless fermions. Firstly, the system is gapped in the bulk. Indeed the low-energy theory of the lattice Hamiltonian Eq.(\ref{KaneMeleHamiltonian}) is directly derived from Eq.(\ref{HaldaneTerm}):
 \begin{equation}
\mathcal{H}_{\rm so} = \Delta_{\rm so} \sigma_3 \tau_3  s_3,
\label{HamiltonianKMEffective}
\end{equation}
where $\Delta_{\rm so}=-3\sqrt{3} t_2$. This perturbation anticommutes with the kinetic Hamiltonian $\mathcal{H}_0$, and therefore opens a gap at the Dirac points. For spinless electrons on the honeycomb lattice, the time-reversal operator is $\mathcal T=\tau_1 K$, where $\tau_1$ switches the valleys and $K$ is the complex conjugation. For spinful electrons it is 
\begin{equation}
\mathcal T=\tau_1 i s_2 K ,
\end{equation} 
where the additional factor $i s_2$ produces the reversal of the electronic spin. It is clear that the mass term $\Delta_{\rm so} \sigma_3 \tau_3  s_3$ is now even under $\mathcal T$ because both $\tau_3$ and $s_3$ change signs under time-reversal. Therefore the spin-coupling coupling allows to open a gap while respecting the fundamental symmetries $\mathcal T$ and $\mathcal P$ of graphene. 
 
\medskip 

{\bf Helical edge mode:} As a spin conserving model, we have seen that the Kane-Mele system consists in two copies of a Chern insulator (or QAH state), each copy being associated with a spin orientation. Since the QAH has a single chiral edge state, we can deduce that the QSH state will have two spin-filtered counter propagating states (Fig. 4).

\medskip

 {\bf Topological robustness in presence of spin mixing:} The intrinsic spin-orbit coupling Eq.(\ref{HamiltonianKMEffective}) leads to the realization of a new state of matter characterized by an insulating gap and a metallic edge. Nevertheless it is quite unlikely that the spin-orbit coupling manifests itself only through this spin-conserving term. Other spin-orbit coupling terms, which mix all spin components, should also be present. Since the previous analysis relies on the description of the system as two decoupled Haldane insulators, it is a natural to ask whether the helical edge states will survive in presence of mixing between the two copies of the Haldane model. Kane and Mele have shown that the counter propagating edge states are in fact robust {\it as long as the bulk is gapped and time-reversal is obeyed}. A Rashba term (due to the presence of a substrate or by a perpendicular electric field) mixes the two spin directions, and spoils the conservation of $s_3$. The corresponding spectrum becomes gapless when the Rashba coupling exceeds the intrinsic spin-orbit coupling \cite{Kane:2005a}. When the Rashba coupling is increased, the bulk gap decreases, but the helical edge states remain gapless (metallic), as long as the gap bulk gap is finite.

\subsection{Experimental realizations}
\label{experimental}
We conclude this section by enumerating some experimental realizations of the topological phases introduced above, including the Quantum Spin Hall (QSH) insulator, the Quantum Anomalous insulator (also called Chern insulator).

\medskip

{\bf The Quantum Spin Hall state:} Unfortunately the QSH state is extremely difficult to observe in graphene due to the actual weakness of the spin-orbit interaction \cite{HuertasPRB:2006,MinPRB:2006}. In 2006, Bernevig, Hughes and Zhang (BHZ) predicted that CdTe/HgTe/CdTe quantum wells should host such a QSH state in their inverted regime \cite{Bernevig:2006}. The transition between the trivial (non inverted regime) and the topological phase (inverted regime) is triggered by varying the thickness of the central HgTe layer of the quantum well. The theoretical prediction was soon followed by the experimental observation of conducting edge states by the group led by Laurens Molenkamp  \cite{Konig:2007}. Using various multi terminal configurations \cite{Roth:2009}, the same group established that their transport measurements are in agreement with counter-propagating edge modes (using a Landauer-Buttiker formalism).

The QSH state has also been predicted in InAs/GaSb quantum wells \cite{CXLiu:2008} with the interesting possibility to tune the transition between the topological and the trivial phases using a gate voltage. The QSH phase and the corresponding edge conduction has been observed in InAs/GaSb quantum wells \cite{Knez:2011}.

\medskip

{\bf The Quantum Anomalous Hall state:}
It has been predicted that the time-reversal invariant QSH state (with its helical edge mode) can be transformed into a QAH state (with its chiral edge mode) by adding magnetic atoms like Mn in CdTe/HgTe/CdTe quantum wells \cite{CXLiu2:2008}. The observation of such a QAH state in magnetic topological insulators has been challenging until it was finally reported in thin films of chromium-doped (Bi,Sb)$_2$Te$_3$ \cite{Chang:2013}, whose Fermi level can be tuned by an electrostatic gate. At zero external magnetic field, the quantization of the Hall resistance was observed at $h/e^2$ in a wide range of gate voltage, with a simultaneous drop of the longitudinal resistance. Those samples are expected to be in the QSH state in the absence of chromium, and chromium apparently develops the suitable spontaneous magnetic order to drive the QSH state into the QHA state. The fact that the longitudinal resistance is not completely vanishing can be explained by the coexistence of the helical and chiral edge modes \cite{Wang:2013}.

\section{Chern insulators: bulk topological invariant}
\label{chern}

In the previous section, we have seen that the Semenov insulator (which breaks inversion symmetry) and the Haldane insulator (which breaks time-reversal symmetry) have the same spectra typical of a massive Dirac fermion. Nevertheless these insulators have very different excitations and physical responses. The Semenov insulator is insulating both in the bulk and along its edge, and has no Hall response. The Haldane insulator is insulating in the bulk, but has also conducting edge channels which carry the integer quantized quantum Hall effect. In this section, we explain that those insulators belong to distinct topological classes of band structures. The Haldane state is characterized by a finite Chern number $C^{-}$ (the upper-script refers to the lowest band) which measures the quantum Hall conductance (in units $e^2/h$) and can be identified to the winding number $n_w$ of a mapping between the BZ and the Bloch sphere (Fig. 5). This winding number is zero for the Semenov insulator.

\subsection{Wavefunctions and Berry phases}
Here we discuss the characterization of topological insulators in terms of the structure of wave funtions. The Hamiltonian Eq. (\ref{2bandH}) describes a generic two band insulator. The band structure consists of an upper ($\alpha=+$) and a lower ($\alpha=-$) bands:
\begin{equation}
E_{\alpha=\pm}(\vk)=\epsilon_0(\vk) \pm |\vd(\vk)|, 
\end{equation}
the corresponding wave functions being the spinors:
\begin{equation}
\Phi^+(\vk) =
\begin{pmatrix}
u_{1}^+ (\vk) \\ 
u_{2}^+ (\vk)
\end{pmatrix}=\begin{pmatrix}
\cos \frac{\theta_\vk}{2} e^{i\phi_\vk}\\ 
\sin \frac{\theta_\vk}{2} 
\end{pmatrix},  
\end{equation}
in the upper band, and 
\begin{equation}
\Phi^- (\vk)=
\begin{pmatrix}
u_{1}^- (\vk) \\ 
u_{2}^- (\vk)
\end{pmatrix}=\begin{pmatrix}
\sin \frac{\theta_\vk}{2} e^{-i\phi_\vk}\\ 
-\cos \frac{\theta_\vk}{2}
\end{pmatrix},  
\label{SpinorPsiMoins}
\end{equation}
in the lower band. The $\vk$-dependent quantities $\theta=\theta_\vk$ and $\phi=\phi_\vk$ are the spherical coordinate angles of the unit vector:
\begin{equation}
\dhat (\vk) = \frac{\vd(\vk)}{|\vd(\vk)|}=
\begin{pmatrix}
\cos \phi_\vk \sin \theta_\vk \\ 
\sin \phi_\vk \sin \theta_\vk \\
\cos \theta_\vk
\end{pmatrix},
\label{dspherique}
\end{equation}
which resides on the unit sphere $S^2$ while $\vk$ spans the $d$-dimensional toroidal Brillouin zone $T^d$. The mapping $\vk \rightarrow \dhat (\vk) = \vd(\vk)/|\vd(\vk)|$ is essential and captures the topological properties of the Hamiltonian $h(\vk)=\vd(\vk) . \boldsymbol{\sigma}$. We assume that the system is insulating so $|\vd(\vk)| \neq 0$ everywhere and the mapping is well-defined over the whole BZ.

We can interpret $\vk$ as a parameter that we can vary along a loop drawn in the BZ and limit ourselves to $d=2$. Along such a loop, the spinor will acquire a Berry phase which is the circulation of the Berry vector potential, also called Berry connection $\mathcal{A}^\alpha (\vk)=(\mathcal{A}^\alpha_x (\vk),\mathcal{A}^\alpha_y (\vk))$ and defined by 
\begin{equation}
\mathcal{A}^\alpha (\vk)= i \sum_{a=1}^{2} (u^\alpha_a)^* \nabla_{\vk} u^\alpha_a,
\label{berry}
\end{equation}
in each band $\alpha=\pm 1$. The Berry curvature is the curl of the Berry connection:
\begin{equation}
F_{xy}^\alpha =[\nabla_\vk \wedge \mathcal{A}^\alpha(\vk)]_z=\partial_{k_x} A^\alpha_y - \partial_{k_y} A^\alpha_x  .
\label{berrycurvature}
\end{equation}
The flux of $\vB_\alpha(\vk)$ through the whole BZ (torus $T^2$), 
\begin{equation}
C^\alpha =\frac{1}{2 \pi} \int_{\rm BZ} d\vk \, F_{xy}^\alpha (\vk),
\label{FluxBerry}
\end{equation}
is called the Chern number of the band $\alpha$. At this stage this $C^\alpha$ is a property characterizing how the spinors wrap or wind around the whole BZ. For most materials, the bands have zero, because a continuous gauge can be defined over the whole BZ and application of the Stokes theorem on a manifold without boundary leads to $C^\alpha=0$. Chern topological insulator (like the Haldane insulator) are systems where it is not possible to define such a unique choice of phases for the spinors \cite{Bernevig:2013}.    
Finally we focus on the lowest band $\alpha=-1$, and evaluate the Berry connection:
\begin{equation}
\mathcal{A}^- (\vk)= i \sum_{a=1}^{2} (u^-)^* \nabla_{\vk} u^-= \sin^2 \frac{\theta_\vk}{2}  \nabla_{\vk} \phi_\vk,
\label{berryconnec}
\end{equation}
and the corresponding Berry curvature:
\begin{equation}
F_{xy}^- =\partial_{k_x} A^-_y - \partial_{k_y} A^-_x =\frac{1}{2} \sin \theta_\vk \left(  \frac{\partial \theta_\vk}{\partial k_x} \frac{\partial \phi_\vk}{\partial k_y} -  \frac{\partial \phi_\vk}{\partial k_x} \frac{\partial \theta_\vk}{\partial k_y}  \right)  .
\label{berrycurvaturemoins}
\end{equation}
The Chern number of the occupied band is given by the formula:
\begin{equation}
C^- =\frac{1}{4 \pi} \int_{\rm BZ} d\vk \,  \sin \theta_\vk \left(  \frac{\partial \theta_\vk}{\partial k_x} \frac{\partial \phi_\vk}{\partial k_y} -  \frac{\partial \phi_\vk}{\partial k_x} \frac{\partial \theta_\vk}{\partial k_y}  \right) ,
\label{FluxBerrymoins}
\end{equation}

\subsection{Hall conductance as a winding number}

The electromagnetic response of 2D two-band insulators can be computed using the Kubo formalism and the expressions of the anomalous current $\boldsymbol{j}$, whose components $j_i$ are defined by ($i=1,2$ refer to $x$- and $y$-axis respectively): 
\begin{equation}
j_{i}=\frac{\partial h(\vk)}{\partial k_i}= \frac{\partial\epsilon_{0}(\vk) }{\partial k_i} \sigma_0 +  \frac{\partial \vd(\vk)}{\partial k_i} .  \boldsymbol{\sigma} .
\label{AnomalousCurrent}
\end{equation}
In space dimension $d=2$, the Hall conductance is exactly given by the Chern number of the lower band. This is also the winding number of the mapping $\vk \rightarrow  \dhat= \vd(\vk)/|\vd(\vk)|$, which explains geometrically the quantization of the Hall conductance in such two-band model. This is reminiscent of the Thouless-Kohmoto-Nightingale-den Nijs (TKKN) invariant for quantum Hall systems \cite{Thouless:1982}. 

The Hall conductivity can be calculated from Kubo formalism as \cite{Qi:2006}:
\begin{equation}
\sigma_{xy} =\frac{e^2}{4 \pi h}  \int d^2 \vk  (f_+(\vk) -  f_-(\vk))   \left(  \frac{ \partial \dhat(\vk)}{\partial k_x}     \times  \frac{ \partial \dhat(\vk)}{\partial k_y}   \right) .   \dhat(\vk) ,
\label{Hallconductance}
\end{equation}
where $f_\pm (\vk)$ are the occupation numbers of the conduction and valence bands. It is assumed that the Fermi level lies in the bulk gap. Hence at zero temperature, where $f_- = 1$ and $f_+ =0$, we have the relation:
\begin{equation}
\sigma_{xy} =\frac{e^2}{h}  n_w ,
\label{Winding1}
\end{equation}
where $n_w$ is the winding number (or Pontryagin index) of the mapping $\vk \rightarrow \dhat (\vk) = \vd(\vk)/|\vd(\vk)|$ between the Brillouin zone (torus $T^2$) and the unit sphere ($S^2$):
\begin{equation}
n_w =\frac{1}{4 \pi}  \int d^2 \vk   \left(  \frac{ \partial \dhat (\vk)}{\partial k_x}     \times  \frac{ \partial \dhat (\vk)}{\partial k_y}   \right) .   \dhat .
\label{Winding2}
\end{equation}
In contrast to the Berry phase, this number is directly constructed from the parameters $\vd(\vk)$ of the Hamiltonian Eq. (\ref{2bandH}) (rather than from derivatives of its eigenstates). This winding number is an integer that counts the number of times the unit vector $\dhat(\vk) $ wraps around the whole sphere $S^2$ while $\vk$ is spanning the whole Brillouin zone $T^2$. In accordance with general classifications, there is a single number that characterizes the general structure of wave functions globally in $\vk$-space. This number is a relative integer, and it measures the charge Hall conductance in units of $e^2/h$. To change $n_w$ it is necessary to change the parameter of the bulk Hamiltonian $\dhat(\vk)$ in such a way that the bulk gap closes. \\

\begin{figure}
\begin{center}
\includegraphics[width=13cm]{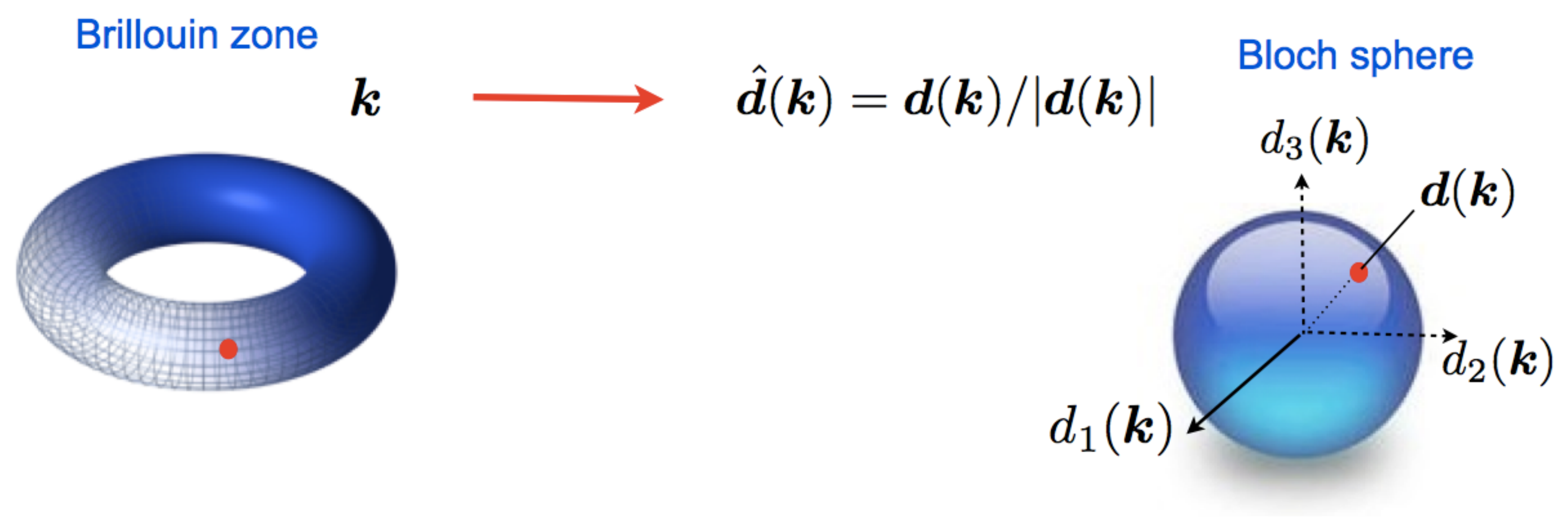}
\caption{Mapping $\vk \rightarrow \dhat (\vk) = \vd(\vk)/|\vd(\vk)|$ between the BZ (torus) and the Bloch sphere. Each point on the unit sphere represents a spinor Eq. (\ref{SpinorPsiMoins}) parametrized by the angles $\theta_\vk$ and $\phi_\vk$.}
\end{center}
\label{FigMapping}
\end{figure}

Let us calculate this winding number for the simple model of massive Dirac fermions introduced previously. We use the parametrization Eq. (\ref{dspherique}) to rewrite the winding number as: 
\begin{equation}
n_w =\frac{1}{4 \pi}  \int d^2 \vk   \sin \theta_\vk \left(  \frac{\partial \theta_\vk}{\partial k_x} \frac{\partial \phi_\vk}{\partial k_y} -  \frac{\partial \phi_\vk}{\partial k_x} \frac{\partial \theta_\vk}{\partial k_y}  \right) = C^-  ,
\label{Winding3}
\end{equation}
which identifies the winding number $n_w$ and the Chern number defined above by Eq. (\ref{FluxBerrymoins}). In principle the number $n_w=C^-$ should be always zero because integration is taken over the whole torus. Finite values can arise from singularities in $ \nabla \phi  $ that are always located at the poles $\theta=0$ and $\theta=\pi$. Hence:
\begin{align}
n_w &= - \frac{1}{4 \pi}  \int d^2 \vk   \nabla_\vk \wedge \left(  \cos \theta_\vk \boldsymbol{\nabla} \phi_\vk   \right) =- \frac{1}{4 \pi}  \int \cos \theta_\vk \boldsymbol{\nabla_\vk} \phi_\vk   . \vd\vl   ,
\label{Winding4}
\end{align}
where the last integral is taken along loops encircling the poles. 

\subsection{Calculations for the Semenov and Haldane insulators}

For graphene, the poles are reached when $\vk$ is at the Dirac points. Then the sign of $d_3(\vk=\pm \vK)$ indicates whether the north or south pole has been reached. For the Semenoff mass, we have $d_3(\vk=\pm \vK) =M_S) $ in both valleys. Then we have to notice that $\vd = (\xi k_x , k_y,M_S) $ accumulates opposite phases, at $\vk=+\vK$ and $\vk=-\vK$, while winding around the same pole (due to the presence of the valley index $\xi=\pm 1$): hence $n_w = 0$. For the Haldane mass, one has $d_3(\vk=\pm \vK) =M_H \xi)$ which means that the accumulated phases (at south and north poles respectively) add up and finally $n_w=1/2+1/2=1$. The general formula is 
\begin{equation}
n_w =- \frac{1}{4 \pi} \sum_{\xi=\pm 1} 2 \pi \xi sign (M_\xi)=\frac{1}{2} \left(  sign(M_-) - sign(M_+) \right),
\end{equation} 
because the winding of the angle $\phi$ in valley $\xi$ is $2 \pi \xi$ and $\cos \theta = Sign (M_\xi)$ where $M_\xi$ is the mass in valley $\xi$. From this formula one sees that the global winding number is zero when the masses are equal in both valleys, and why it is $n_w =\pm 1$ in the Haldane phase characterized by a band inversion. This has been formulated in a more elegant and general way in Ref. \cite{Sticlet:2012}.

\medskip

Let us now consider the example of graphene in presence of some inversion breaking and time-reversal breaking terms. So the mass matrix is $(M_S - 3 \sqrt{3} t_2 \sin (\phi) \xi ) \sigma_3$ implying that the gap can close for $M_S =\xi 3 \sqrt{3} t_2 \sin \phi$ in one valley (labelled by $\xi=\pm 1$). This equality signals a one-electron topological quantum transition separating a QHA insulator and a trivial atomic insulator. Finally we would like to make a comment on the terminology. This type of phase transition is purely a change between two one-electron Hamiltonians. It has in particular nothing to do with topological order defined by Wen. In particular the transition discussed here is not a transition between two topological orders. It is rather a transition between two band-insulators having distinct topological invariants (which characterize the winding of one-electron wave functions).

\section{Edge states}
\label{SectionEdge}
The 2D topological insulators (the QAH and the QSH states) are insulating in the bulk and conducting along their edge. They are characterized by:
\begin{itemize} 
\item a bulk topological invariant (the Chern number for QAH, and the Z$_2$ index for QSH insulators) 
\item characteristic edge states (chiral for QAH, and helical for QSH states)
\end{itemize}
Here we describe more thoroughly this deep connection between bulk band structure properties and the existence of edge states by computing explicitly the edge state running at the interface between different pairs of insulating phases. Two different situations are to be contrasted. First one might consider a mass kink without change of the topological invariant, for instance an interface between two Semenov insulators characterized by opposite values of the parameter $M_S$, or two Haldane insulators (in the terminology introduced in section \ref{sectionGraphene}) having opposite masses. Then the edge states exist but they are not protected against scattering. In contrast, the edge states residing at the interface between a Semenov phase and a Haldane phase (two topologically distinct insulators) are topologically protected, and remain metallic as long as the bulk gaps are preserved. Our examples are specialized in spacial dimension 2, implying that the edge modes are running along 1D interfaces. Nevertheless the idea is rather general and valid for $D-1$ dimensional surface gapless modes emerging at interfaces between $D$ dimensional gapped phases. This physics is reminiscent of the Jackiw-Rebbi model introduced in field theory \cite{Jackiw:1976} and of the physics of solitons in polyacetylene (D=1) \cite{SSH:1979,SSH:1980}.  

\subsection{Interface between topologically distinct insulators \label{HaldaneSemenov}}

\begin{figure}
\begin{center}

\includegraphics[width=9cm]{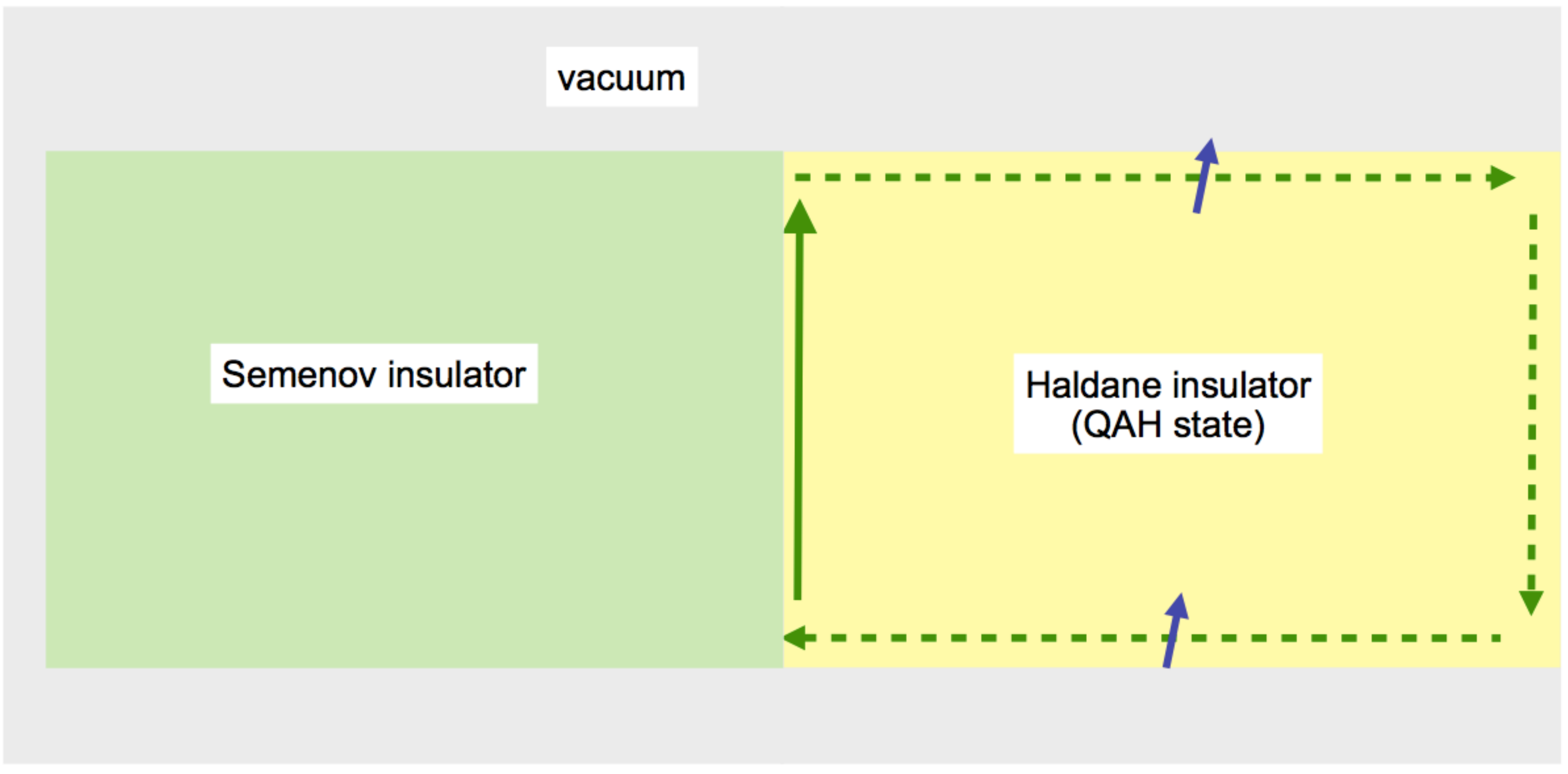}
\hspace{1 cm}
\includegraphics[width=6cm]{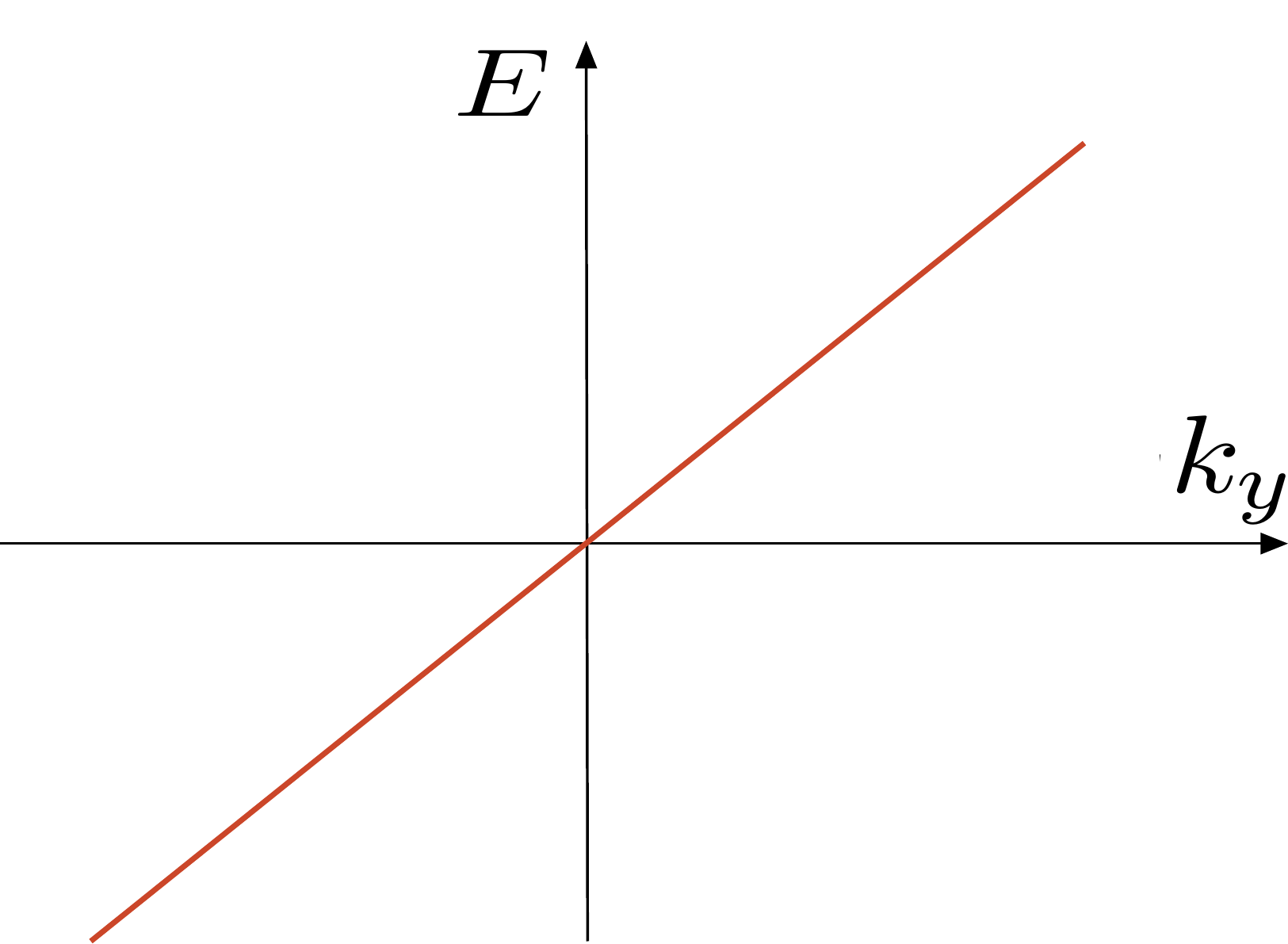}
\caption{{\it Left panel:} Interface between the Semenov insulator ($x<0$, light green) and the Haldane insulator ($x>0$, yellow). There is always a zero-energy bound state located near the interface $x=0$. {\it Right panel:} dispersion $E(k_y)={\rm sign}(M_H) \hbar v_F k_y $ of the chiral edge mode along the $y-$axis (solid green arrow).  }
\end{center}
\label{FigHaldaneSemenovWall}
\end{figure}

We assume that one half-plane ($x<0$) is filled with a "Semenov insulator" while an "Haldane insulator" occupies the other half-plane ($x>0$). In principle one should define this heterojunction on the lattice by varying the mass parameters of the model (namely the on-site mass $M_S$ and the chiral phase $\phi$) near the interface. Since we are mainly interested in eventual zero modes confined near the interface $x=0$, we use the low energy effective model valid near the Dirac points (energies smaller than the bandwidth $t$).  Using the translational invariance along the $y-$direction, the wave function can be written $\Psi(\vr)=\Psi(x) e^{i k_y y}$ and the corresponding wave equation for $\Psi(x)$ reads: 
\begin{equation}
\left( -i \hbar v_F \sigma_1 \tau_3 \partial_x +\hbar v_F k_y  \sigma_2 + M(x)  \right) \Psi(x) = E \Psi(x),
\label{waveeq}
\end{equation}
with $M(x)=M_S \Theta(-x) \sigma_3 + M_H \Theta(x) \sigma_3 \tau_3$. In fact we can even consider a more general shape by replacing the Heaviside functions $\Theta(x)$ with smooth functions interpolating between zero for negative arguments and one for large positive arguments. Nevertheless the sharp step model is accurate, provided the length scale for the variation of the lattice parameters is smaller than the extension of the eventual edge state, namely $\hbar v_F /{\rm max}(M_{S},M_{H})$. The two sets of Pauli matrices $\sigma_i$ and $\tau_i$ represent respectively the sublattice isospin and the valley degrees of freedom.   

\medskip

One first shows the existence of a zero energy solution at $k_y=0$, by solving the equation:
\begin{equation}
\hbar v_F \partial_x \Psi(x) = -i \sigma_1 \tau_3  M(x) \Psi(x) ,
\end{equation}
obtained by multiplying both sides of Eq. (\ref{waveeq}) by $i \sigma_1 \tau_3$.
In the region $x>0$, this equation reads:
\begin{equation}
\hbar v_F \partial_x \Psi = - \sigma_2   M_H \Psi ,
\end{equation}
and the bounded solution (decaying at $x \rightarrow \infty$) is the eigenstate of $\sigma_2$ with eigenvalue sign$(M_{H})$

For $x<0$, there is an additional valley matrix $\tau_3$ in the wave equation:
\begin{equation}
\hbar v_F \partial_x \Psi = - \sigma_2 \tau_3   M_S \Psi ,
\end{equation}
and the corresponding bounded solution is the eigenstate of $\sigma_2$ with the eigenvalue: -sign$(\xi M_S)$. So the matching is possible, and there is a zero mode at the boundary ($x=0$), only if the two solutions above correspond to the same eigenvalue of $\sigma_2$, namely if
\begin{equation}
sign(M_{H})=-sign(\xi M_S).
\end{equation}
For any choice of the masses, this equality is always valid in one valley which is fixed by the relative signs of $M_{H}$ and $M_{S}$. Therefore one always obtains a zero mode which {\it is polarized in the valley $\xi=-sign(M_{S}M_{H})$}. 

\medskip

Now in order to obtain the wave function and dispersion $E(k_{y})$ of this edge mode, let us restore finite energy $E$ and parallel momentum $k_y$ in Eq.(\ref{waveeq}). Without any further calculation, one simply notices that the zero mode at $k_y=0$ is also eigenstate of $\hbar v_F k_y  \sigma_2$, and therefore the expression of its wave function is still valid at finite energy and momentum with the dispersion:
\begin{equation}
E=-sign(\xi M_{S}) \hbar v_F k_y =sign(M_{H}) \hbar v_F k_y .
\end{equation}
The edge mode is chiral and shows up in the valley that is experiencing a mass inversion at the interface. In the limit of large $M_S$, the Semenov insulator can represent the vacuum. By reproducing this calculation for various orientation of the interface it is easy to demonstrate that the Haldane insulator is surrounded by a 1D edge chiral edge mode that circulates clockwise if $sign(M_S M_H)$ is positive, and anti-clowise for negative $M_S M_H$. Note that if we assume that the vacuum is represented by a large positive $M_S$, then the sign of $M_S M_H$ is simply the sign of $M_H=-3\sqrt{3} t_2 \sin(\phi)$ which is set by the chirality of the flux pattern in the microscopic Haldane model.


\subsection{Kink in the Haldane mass}
We consider now a linear junction between two Haldane insulators with opposite chiralities. In the low-energy effective model, the full wave equation for this situation reads: 
\begin{figure}
\begin{center}
\includegraphics[width=9cm]{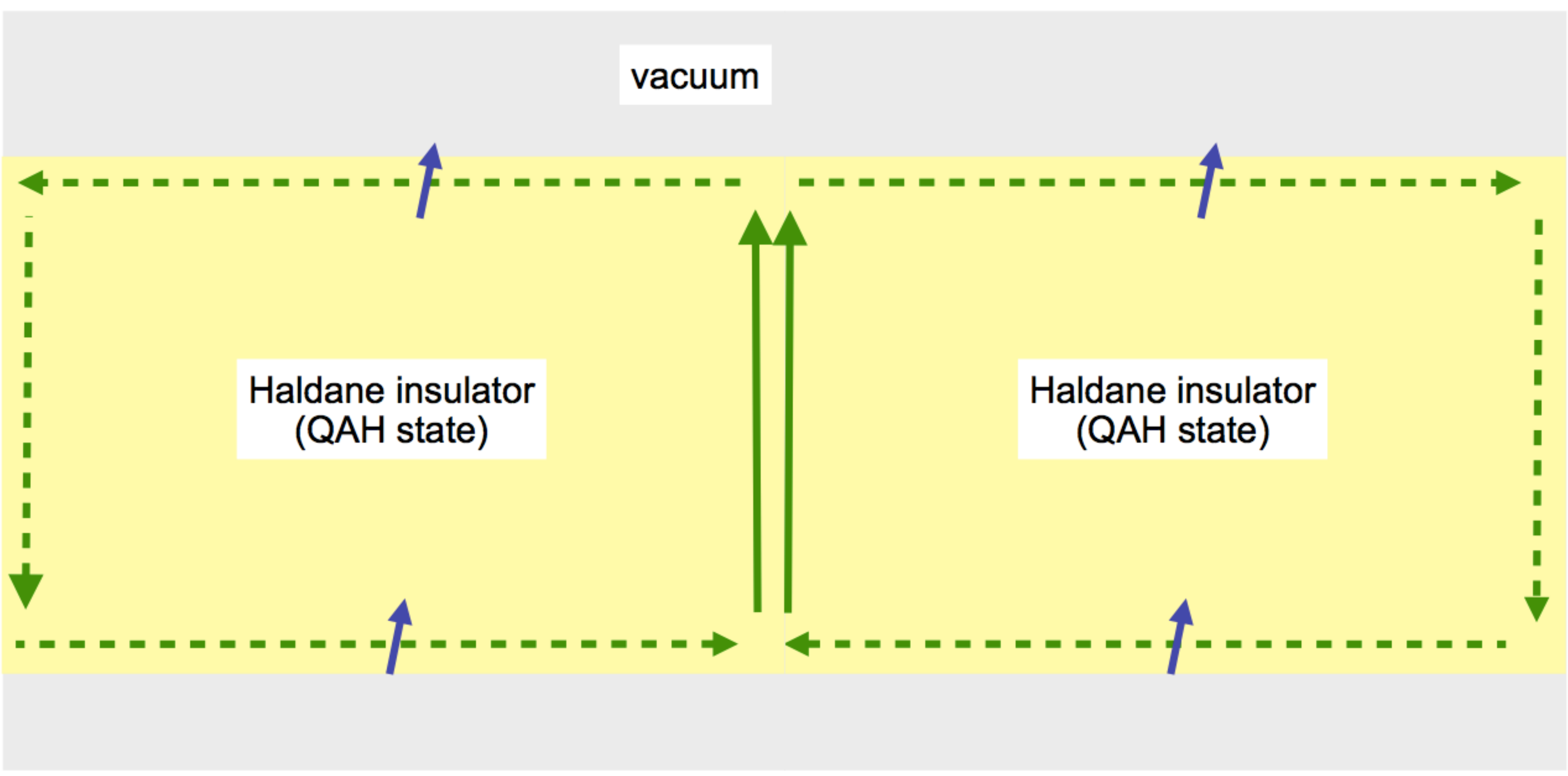}
\hspace{1 cm}
\includegraphics[width=6cm]{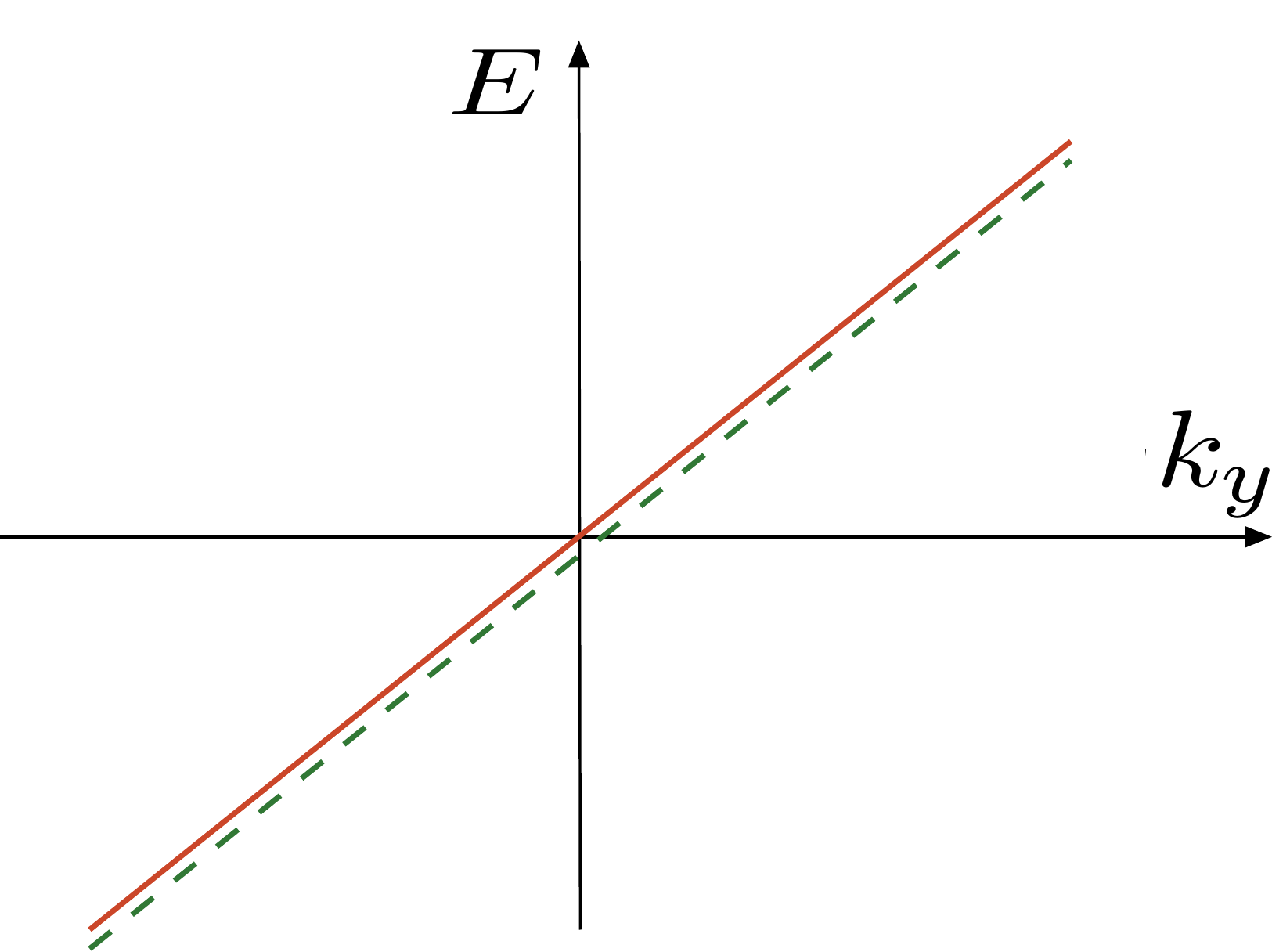}
\caption{{\it Left panel:} Kink in the Haldane mass with the sign change occurring at $x=0$. There is always twofold degenerate zero-energy bound states located near $x=0$ (indicated by 2 solid arrows in green). {\it Right panel:} dispersion $E(k_y)={\rm sign}(M_H) \hbar v_F k_y $ of the two independent chiral edge modes (solid and dashed curves respectively).}
\end{center}
\label{FigKinkHaldane}
\end{figure}
\begin{equation}
\left( -i \hbar v_F \sigma_1 \tau_3 \partial_x +\hbar v_F k_y  \sigma_2 + M_H(x) \sigma_3 \tau_3  \right) \Psi(x) = E \Psi(x),
\label{waveeqKinkHaldane}
\end{equation}
where $M_H(x)$ is a real monotonic function describing a kink with $M_H(\infty)$ positive and $M_H(-\infty)$ negative hereafter (the other kink configuration can be treated similarly). We choose the origin ($x=0$) where $M(x)$ has its zero. We expect that a bound state might show up near $x=0$ because the insulator becomes "locally" gapless there. 

We first look for a $E=0$ solution at $k_y=0$ by solving the equation: 
\begin{equation}
 i \hbar v_F \sigma_1 \tau_3 \partial_x \Psi(x)  = M_H(x) \sigma_3 \tau_3  \Psi(x) .
\label{waveeqKinkHaldane2}
\end{equation} 
By multiplying each side by $-i \sigma_1 \tau_3$, it is obtained:
\begin{equation}
\hbar v_F  \partial_x \Psi(x)  = - M_H(x) \sigma_2 \tau_0 \Psi(x) ,
\label{waveeqKinkHaldane3}
\end{equation} 
which has the solution (valid for all values of $x$):
\begin{align}
\Psi(x) & =\exp \left( - \int_{0}^{x} dx' M_H(x')/\hbar v_F \right) | \sigma_2 \tau_0=+1 \rangle \\
 &= \exp \left( - \int_{0}^{x} dx' M_H(x')/\hbar v_F \right) 
 \left[
a
\begin{pmatrix}
1 \\ 
i\\
0\\
0\\
\end{pmatrix}
+
b
\begin{pmatrix}
0 \\ 
0\\
1\\
i\\
\end{pmatrix}
\right],
\label{waveeqKinkHaldaneSolution}
\end{align} 
where $a$ and $b$ are complex coefficients. Hence there is a twofold degenerate zero mode (at $k_y=0$) due to the presence of the identity matrix $\tau_0$ in Eq. (\ref{waveeqKinkHaldane3}).

\medskip

Now we can restore a finite transverse momentum $k_y$ and observe that the above solution is an eigenmode of $\hbar v_F \sigma_2$ with energy $E=\hbar v_F k_y$. The two degenerate chiral zero modes Eq.(\ref{waveeqKinkHaldaneSolution}) yield two degenerate chiral modes propagating in the same direction along $y$-axis:
\begin{equation}
\Psi(\vr)=\Psi(x) e^{i k_y y} =\exp \left( i k_y y  - \int_{0}^{x} dx' M_H(x')/\hbar v_F  \right) | \sigma_2 \tau_0=+1 \rangle.
\label{waveeqKinkHaldaneSolutionFinite}
\end{equation} 
This is consistent with the fact that the Haldane model breaks time-reversal symmetry. We can understand the Haldane kink as two remote Haldane insulators (with opposite chiralities) that would have been brought in contact adiabatically. After such a process one would have two modes running in the same direction along the interface considered.

\subsection{Kink in the Semenov mass}

\begin{figure}
\includegraphics[width=9cm]{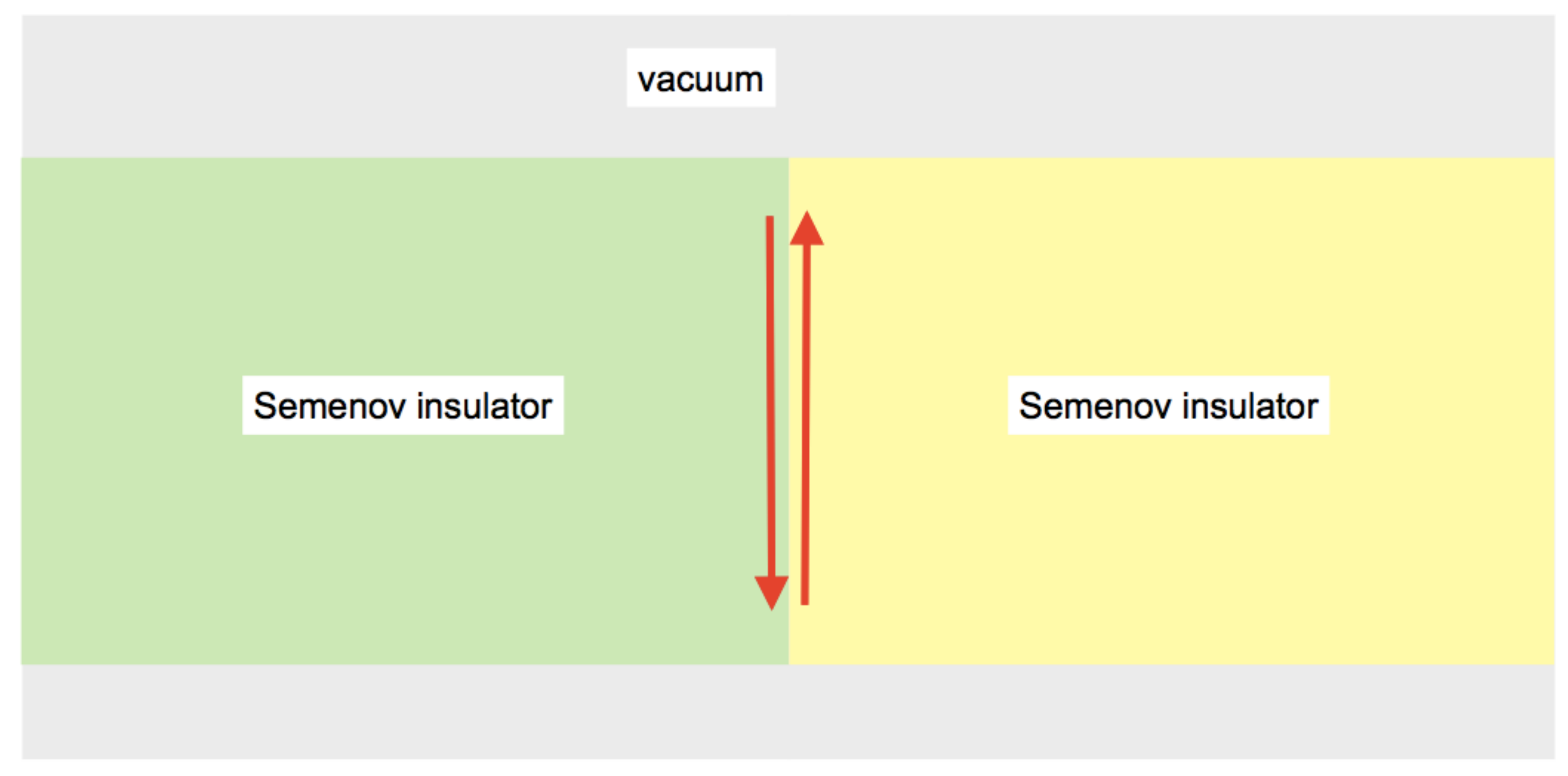}
\hspace{1 cm}
\includegraphics[width=6cm]{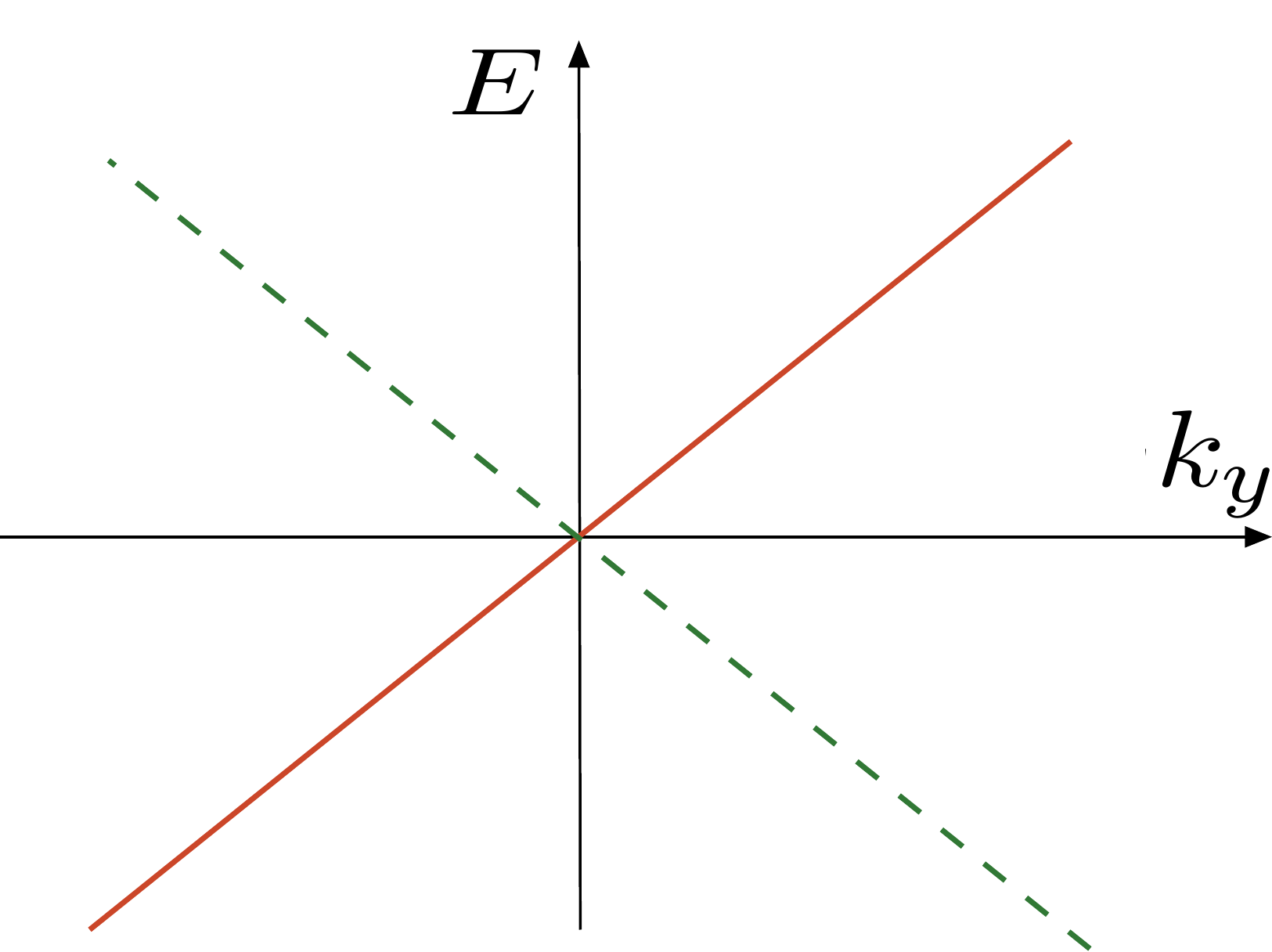}
\caption{{\it Left panel:} Kink in the Semenov mass $M_S(x)=M(x) \sigma_3$ with the sign change occurring at $x=0$. There is always twofold degenerated zero energy bound states located near $x=0$ (indicated by the solid lines in red). {\it Right panel:} dispersion $E(k_y)=\pm \hbar v_F k_y $ of the two counter-propagating edge modes (solid and dashed curves respectively). This counter propagation is the natural consequence of the time-reversal invariance of the system.}

\label{FigKinkSemenov}
\end{figure}

One can easily reproduce the similar analysis for a kink of the Semenov mass by solving the wave equation: 
\begin{equation}
\left( -i \hbar v_F \sigma_1 \tau_3 \partial_x +\hbar v_F k_y  \sigma_2 + M_S(x) \sigma_3  \right) \Psi(x) = E \Psi(x),
\label{waveeqKinkSemenov1}
\end{equation}
where $M_S(x)$ is a real function satisfying $M_S(0)=0$, $M_S(\infty)>0$ and $M_S(-\infty)<0$. The equation for the eventual zero energy mode at $k_y=0$ is then:
\begin{equation}
\hbar v_F  \partial_x \Psi  = -M_S(x)  \sigma_2 \tau_3 \Psi ,
\label{waveeqKinkSemenov2}
\end{equation} 
whose solution reads:
\begin{align}
\Psi(x) & =\exp \left(- \int_{0}^{x} dx' M_S(x')/\hbar v_F \right) | \sigma_2 \tau_3=+1 \rangle \\
 &= \exp \left( - \int_{0}^{x} dx' M_S(x')/\hbar v_F \right) 
 \left(
a
\begin{pmatrix}
1 \\ 
i\\
0\\
0\\
\end{pmatrix}
+
b
\begin{pmatrix}
0 \\ 
0\\
1\\
-i\\
\end{pmatrix}
\right),
\label{waveeqKinkSemenovSolution}
\end{align}
where $a$ and $b$ are complex numbers. As  a major difference with the Haldane kink, the two parts of the wave function leads to opposite chiralities when a finite $k_y$ is restored. This is because they correspond to eigenmodes of $\sigma_2$ with opposite eigenvalues $\pm 1$. This is consistent with the global time-reversal symmetry of the system.

\section{Conclusion}

In high-energy physics, spin one-half fermions are described by fields whose free dynamics follow the wave equations initially discovered by P. Dirac in 1928 \cite{Dirac:1928}, and H. Weyl in 1929 \cite{Weyl:1929} (Sec. \ref{SectionWave}). In "Dirac materials", the Bloch wave functions follow Dirac-like or Weyl-like equations at least at vicinity of some special points of the Brillouin zone. As a famous example of a "Dirac material', semimetallic graphene hosts low-energy excitations behaving as massless fermions described by a 2D Weyl equation near two isolated points of the BZ (Sec. \ref{SectionGraphene}). Some insulators with strongly coupled bands are locally (in $\vk$-space) described by a Dirac equation where the mass is replaced by the band gap. In condensed matter systems, the discrete degrees of freedom that couple to the quasi-momentum $\vk$ might be related to many origins: real electronic spin, sublattice isospin, orbital index, etc... Therefore one might obtain different kinds of insulting states, like the Semenov, the Haldane or the Kane-Mele insulators reviewed in Sec. \ref{insulators}.

 In contrast to experiments in high-energy colliders, electrons in materials (or cold atoms in optical lattices) are affected by the presence of a dense lattice. As a consequence, Bloch shown that the electronic spectrum form bands $E(\vp)$ that are periodic in the quasi momentum $\vp$, i.e. the reciprocal space becomes a compact manifold: the Brillouin zone (BZ). This implies the possibility of some nontrivial wrapping of the wave functions around the BZ, which leads to topological properties like the quantization of the integer quantum Hall effect (Sec. \ref{chern}) and the existence of protected edge states in some insulators (Sec. \ref{SectionEdge}).  

\medskip

Finally, condensed matter and cold atom systems allow the interesting possibility of modifying the topological properties of Dirac insulators by applying time-dependent perturbations to the system. For instance, an ordinary insulator or semimetallic graphene could be driven into the Haldane insulating phase by applying circularly polarized light \cite{Oka:2009,Kitagawa:2011,GuPRL:2011,torres}, or into a QSH state by applying the suitable linearly polarized light \cite{lindner,Cayssol:2013}. Another exciting perspective is the recent realizations of artificial honeycomb lattice systems that mimics the graphene. Those "artificial graphenes" (see Ref. \cite{Polini:2013} for a review) are realized with cold atoms loaded in optical lattices \cite{tarruell2011}, with molecules deposited on a metallic surface by a STM tip \cite{gomes2011}, or by nanopatterning a two-dimensional electron gas \cite{West,Park}.




\section*{Acknowledgements}
I would like to thank Doru Sticlet for discussions and for his careful final reading of the manuscript.
Besides it is a pleasure to thank all my colleagues at Berkeley and in Dresden (Max Planck Institute for Complex Systems) for their hospitality during my 
Marie-Curie mobility, and for invaluable discussions with them. I acknowledge support from EU/FP7 under contract TEMSSOC and from the french Agence Nationale de la Recherche through project
2010-BLANC-041902 (ISOTOP).

\bibliographystyle{unsrt}

\bibliography{HdrThesis}




\end{document}